\documentclass[a4paper,11pt]{article}

\pdfoutput=1 

\usepackage{jheppub} 
                    
\usepackage{arydshln}
\usepackage[T1]{fontenc} 

\usepackage{amsmath}    
\usepackage{graphicx}   
\usepackage{subfigure}
\usepackage{color}

\usepackage{bm} 
\usepackage{soul}
\usepackage{graphicx}
\usepackage{amsmath,amssymb}
\usepackage{soul}
\usepackage{textcomp}
\usepackage{hyperref}
\usepackage{subfigure}
\usepackage[toc,page]{appendix}
\usepackage{ulem}

\graphicspath{{./fig/}}
\def \pd {\partial}

\def\st{\begin{equation}}
\def\stp{\end{equation}}
\def\bg{\begin{eqnarray}}
\def\nd{\end{eqnarray}}

\def \bes {\begin{subequations}}
\def \ees {\end{subequations}}

\def \a {\alpha}
\def \b {\beta}

\def \d {\delta}
\def \e {\epsilon}

\def \o {\omega}

\def \G {\Gamma}

\def \vx {\bm{x}}
\def \vy {\bm{y}}

\def \vQ {\bm{Q}}

\def \<{\langle}
\def \>{\rangle}
\def \+{\dagger}
\def \({\left(}
\def \){\right)}
\def \[{\left[}
\def \]{\right]}
\def\no{\nonumber}

\def\crit{\textrm{crit}}
\def\Ising {\textrm{Is}}

\def\QGP{\textrm{QGP}}
\def\CO{\textrm{C.O.}}
\def\noCP{\textrm{no C.P.}}

\def\eff{\textrm{eff}}
\def \fm {\textrm{fm}}

\def \GeV {\textrm{GeV}}
\def\BR{\textrm{B.R.}}
\def\noBR{\textrm{no B.R.}}

\def\plus{(+)}
\def\phieq{\overline{\phi}}
\def\Gammaeq{\overline{\G}}

\def \QKZ {Q^{*}}

\def \Qscaled{\widetilde{Q}}

\definecolor{dgreen}{rgb}{0.0, 0.5, 0.0}

\begin{document}

\title{Hydro+ in Action: Understanding the Out-of-Equilibrium Dynamics Near a 
Critical Point in the QCD Phase Diagram
}

\vspace{5mm}
\preprint{MIT-CTP-5142}

\author[a]{Krishna Rajagopal}
\author[a]{Gregory Ridgway}
\author[a]{Ryan Weller}
\author[a]{Yi Yin}

\affiliation[a]{Center for Theoretical Physics, Massachusetts Institute of Technology, Cambridge, Massachusetts 02139, USA }

\emailAdd{krishna@mit.edu}
\emailAdd{gregridgway@gmail.com}
\emailAdd{rweller@mit.edu}
\emailAdd{yiyin3@mit.edu}


   \abstract{
   Upcoming experimental programs, including the Beam Energy Scan at RHIC, 
   will look for signatures of a possible critical point in the QCD phase diagram in fluctuation observables.  
   To understand and predict these signatures, one must account for the fact that the dynamics of any critical fluctuations must be out-of-equilibrium: 
   because of critical slowing down, the fluctuations cannot stay in equilibrium as the droplet of QGP produced in a collision expands and cools. 
   Furthermore, their out-of-equilibrium dynamics must also influence the hydrodynamic evolution of the cooling droplet.  
   The recently developed Hydro+ formalism allows for a consistent description of both the hydrodynamics and the out-of-equilibrium fluctuations, 
   including the feedback between them. 
   We shall provide an explicit demonstration of how this works, 
   setting up a Hydro+ simulation in a simplified setting: 
   a rapidity-independent fireball undergoing radial flow with an equation of state in which we imagine a critical point close to the $\mu_B=0$ axis of the phase diagram. 
   Within this setup, we show that we can quantitatively capture non-equilibrium phenomena, including critical fluctuations over a range of scales and memory effects. Furthermore,  we illustrate the interplay between the dynamics of the fluctuations and the hydrodynamic flow of the fireball: as the fluid cools and flows, the dynamical fluctuations lag relative to how they would evolve if they stayed in equilibrium; there is then a backreaction on the flow itself due to the  out-of-equilibrium fluctuations; and, in addition, the radial flow transports fluctuations outwards by advection.
Within our model, we find that the backreaction from the
out-of-equilibrium fluctuations
does not yield dramatically large effects in the hydrodynamic variables. Further work will be needed in order to check this quantitative conclusion in other settings but, if it persists, this will considerably simplify future modelling.
   }
   

\date{\today}
\maketitle

\section{Introduction}
\label{sec:intro}

Does the rapid but smooth crossover between hadron gas and quark gluon plasma (QGP) at small baryon chemical potential $\mu_{B}$~\cite{Aoki:2006we,Bazavov:2009zn,Borsanyi:2010cj,Bazavov:2011nk} 
turn into a first order phase transition beyond some critical point at a nonzero $\mu_{B}$?
This is one of the main unanswered 
questions about the phase structure of QCD matter~\cite{Berges:1998rc,Halasz:1998qr,Stephanov:1998dy,Stephanov:1999zu,Rajagopal:2000wf,Stephanov:2007fk,Fukushima:2010bq,Luo:2017faz,Busza:2018rrf,Bzdak:2019pkr}.
To date, because of the fermion sign problem {\it ab initio} lattice calculations have proved 
prohibitively challenging at nonzero $\mu_B$ except via Taylor expansion in $\mu_B/T$ about $\mu_B=0$, analytic continuation from imaginary $\mu_B$, or via reweighting, all of which are algorithms that are based upon extracting physics at nonzero $\mu_B$ from lattice calculations whose intrinsic formulation is at $\mu_B=0$.
In contrast, because the ions that collide in heavy ion collisions
carry nonzero net baryon number, the QGP produced in these collisions
naturally inherits some nonzero $\mu_B$ -- it is produced doped with baryons.  At the highest collision energies accessible at the Relativistic Heavy Ion Collider (RHIC) and even more so at the Large Hadron Collider (LHC) most of the net baryon number
coming from the incident nuclei ends up at high rapidity and the
QGP at mid-rapidity is almost undoped.  Studying the properties
of QGP as a function of its doping with baryon number, which is to say mapping the phase diagram of QCD at nonzero $\mu_B$, requires
analyzing heavy ion collisions at lower collision energies
where more net baryon number ends up at mid-rapidity and QGP doped
to a larger $\mu_B$ is produced.
This is the goal of the
ongoing Beam Energy Scan (BES) program at RHIC.
Data taking in the second, high statistics, phase of this program
began in 2019 and is expected to conclude in 2021.
The BES program
provides a unique opportunity to detect signatures
of the QCD critical point if such a point exists within the region of the phase diagram that is accessible to experiment.
This program relies upon the
Low Energy RHIC electron Cooling (LEReC) upgrade to the RHIC accelerator which will increase its luminosity at low
energies as well as
upgrades to the STAR detector --- the Inner Time Projection Chamber (iTPC), Event Plane Detector (EPD) and Endcap Time of Flight detector (EToF).
These enhancements to RHIC and STAR have been designed, in concert, to
bring out fluctuation
observables sensitive to the presence of a possible 
critical point with unprecedented high statistics 
for collisions with
center-of-mass energies ranging between $7.7~\GeV$ per nucleon and $20~\GeV$ per nucleon, producing droplets of QGP that freezeout with $400~$~MeV$\gtrsim\mu_B\gtrsim 200$~MeV.
STAR also plans to do fixed-target collisions that will extend the reach of this program to lower center-of-mass energies,
producing QGP at even larger $\mu_B$.
In addition to the BES program at RHIC,
there are a number of other approved experiments anticipated in the coming years, 
including the Compressed Baryonic Matter (CBM) experiment at the FAIR facility at GSI, the Multi-Purpose-Detector (MPD) at the NICA accelerator in Dubna,  and the CSR-External target Experiment (CEE) at the HIAF facility in China.

The anticipated experimental advances motivate a substantial
theoretical effort needed to meet the challenges involved in identifying
signatures of possible critical fluctuations. 
To maximize the discovery potential of the experimental efforts, 
it is crucial to understand signatures originating from
the fluctuation of the critical order parameter field.
In thermal equilibrium, such fluctuations grow according to
universal scaling laws the closer one
gets to a critical point, for example as a function of increasing
$\mu_B$, and then decrease again as the critical point is passed.
In particular, therefore, we expect a nonmonotonic dependence on $\mu_B$, and hence collision energy, in the
fluctuations of the measured multiplicity of various species of hadrons~\cite{Stephanov:1998dy,Stephanov:1999zu}, 
most notably the non-Gaussian fluctuations in the multiplicity of protons~\cite{Hatta:2003wn,Stephanov:2008qz,Athanasiou:2010kw,Luo:2017faz}. Because $\mu_B$ also depends on rapidity, it is also
of interest to analyze the rapidity dependence of these observables in collisions with a given energy~\cite{Brewer:2018abr,Shen:2018pty}
Furthermore,  
the enhancement of critical fluctuation would induce 
universal singular behavior in the Equation of State (EoS) and in particular in transport coefficients such as bulk viscosity at and around the hypothesized critical point.
See Ref.~\cite{Parotto:2018pwx} for 
the construction of a family of EoS that incorporate 
the expected critical behavior as well as what is known about QCD thermodynamics at lower $\mu_B$ from
lattice calculations, and see Refs.~\cite{Monnai:2016kud,Martinez:2019bsn} for discussion of the behavior of bulk viscosity near the critical point. 
Since the EoS and transport coefficients control 
the bulk evolution of the QGP droplet produced in a heavy ion collision, signatures of a possible
critical point could also manifest themselves in observables which reflect the characteristics of hydrodynamic evolution.  These, too, are a focus of the BES program.

However, 
it has long been understood~\cite{Berdnikov:1999ph} that
critical fluctuations cannot possibly stay in thermal equilibrium
during a heavy ion collision.  
If there is a critical point in the equilibrium
phase diagram of QCD the droplet of QGP formed in a heavy
ion collision may indeed pass near it as it expands and cools. 
But, because the time that these rapidly cooling droplets of 
hot matter spend in the vicinity of the critical
point is finite, and because 
long wavelength critical fluctuations 
are intrinsically slow to equilibrate, with the equilibration timescale diverging near the critical point (a phenomenon called critical slowing down),
the critical fluctuations inescapably fall out-of-equilibrium. 
As was already apparent in the earliest work~\cite{Berdnikov:1999ph} and has been much further understood more recently~\cite{Mukherjee:2015swa} (see Ref.~\cite{Yin:2018ejt} for a brief review)
the out-of-equilibrium fluctuations can be 
quite different from equilibrium expectations, certainly quantitatively and even qualitatively. 
Furthermore, 
these out-of-equilibrium fluctuations must also 
modify the equation of state (EoS), changing it from
what it would be in equilibrium\footnote{To see why this must be so, let us start by recalling that in equilibrium the correlation length of the critical fluctuations diverges and this affects the EoS, for example causing the specific heat to have a singularity. Critical slowing down means that in reality the fluctuations do not stay in equilibrium and in particular their correlation length does not diverge. This means that the specific heat should not be expected to have a singularity.  This is just one example of how the out-of-equilibrium fluctuations must modify the EoS relative to what it would have been if the fluctuations were able to stay in equilibrium.}, 
and this means that
they must influence the hydrodynamic evolution.
The hydrodynamic evolution (expansion and cooling) 
drives the fluctuations away from equilibrium, and through the EoS this
in turn must modify the hydrodynamic evolution.

We will adopt the newly developed Hydro+ framework to study the
 intertwined dynamics between the evolution of out-of-equilibrium fluctuations and the bulk hydrodynamic evolution~\cite{Stephanov:2017ghc}. 
In this approach,
the dynamics of hydrodynamic variables as well as the long wavelength critical fluctuations 
are studied self-consistently by solving a set of coupled deterministic equations.
The Hydro+ approach can describe nontrivial critical dynamics including critical slowing down, 
as well as modifications of the sound velocity and bulk viscosity (relative to their equilibrium values) caused by out-of-equilibrium fluctuations~\cite{Stephanov:2017ghc}. Via these modifications,
there is a feedback on the bulk flow coming from the out-of-equilibrium critical fluctuations.

Quantitative and self-consistent modelling 
of the nonequilibrium evolution of critical fluctuations in heavy ion collisions is a central challenge for theoretical physicists at the present time, with the RHIC BES program now underway.  This is
a core goal of the Beam Energy Scan Theory collaboration, and in
its full form this will require understanding of and controlled modelling of the initial stages of the collision and the freezeout dynamics as well as the coupled evolution of hydrodynamics and
critical fluctuations that occurs between the early and late
stages of the collision and that we shall consider.  Without a treatment of the early and late stages of the collision, our study cannot by itself yield predictions to be compared to experiment.
However, having the means to follow the out-of-equilibrium
evolution of critical fluctuations, quantitatively, in a way
that incorporates their influence on the bulk evolution as
well as the influence of the bulk evolution on them
is a necessary ingredient to any future effort to
extract information about the presence and location
of a critical point from experimental data.

The Hydro+ formalism is built upon deterministic equations 
for two-point functions (and, in future, higher-point functions)
of the fluctuations. This 
is not the only possible formalism with which
to achieve our goals.
See Refs.~\cite{Kapusta:2012zb,Kapusta:2017hfi,Sakaida:2017rtj,Nahrgang:2018afz} for studies using complementary approaches based upon simulating stochastic equations.  
See also Refs.~\cite{Akamatsu:2016llw,Murase:2016rhl,Hirano:2018diu,Singh:2018dpk} for related developments in the context of fluctuating hydrodynamics away from a critical point
and other studies of non-equilibrium effects around the phase transition found in Refs.~\cite{Nahrgang:2011mv,Nahrgang:2011mg,Herold:2016uvv,Herold:2018ptm}, which include studies of the back reaction of
the order parameter fluctuations on the (stochastic) hydrodynamic variables.

The quantitative description of the future BESII data requires inputting realistic initial conditions at the relative low beam energy, solving 3d hydro+ equation at finite baryon density, and doing the appropriate freezeout of critical fluctuations.
We shall not do such a study here.

%
%
%
\begin{figure} 
\center
\includegraphics[width=0.65\textwidth]{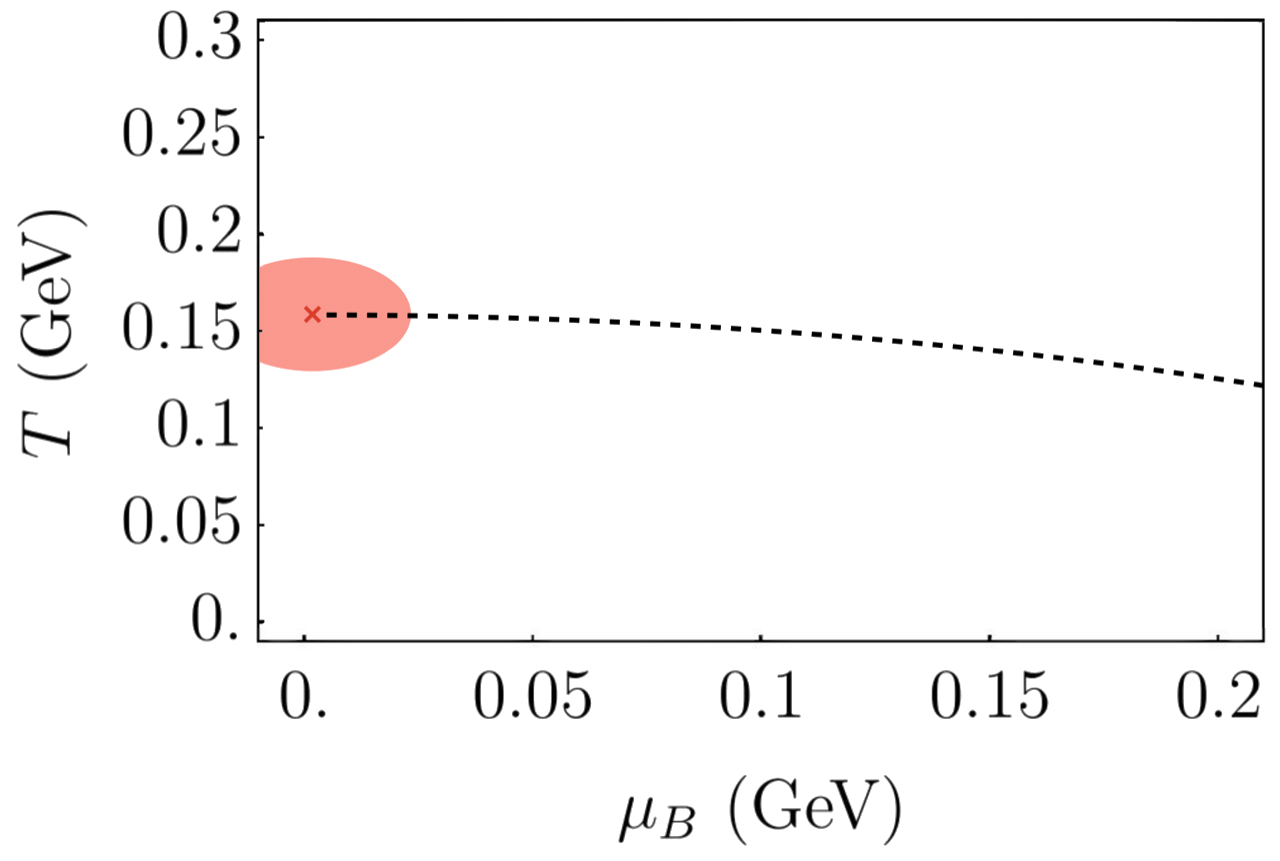}
\caption{
\label{fig:critical_point}
Illustration of the phase diagram with a critical point (red x) that we imagine, surrounded by a pink region that gives an impression of the critical region within which the dynamics of critical fluctuations are important. For example, the boundary of the critical region could be the curve inside which the equilibrium correlation length is greater than 0.5 fm. With any such definition, the critical region will have a nontrivial shape; it will not be just an ellipse as in our illustration.  
To simplify our analysis we have imagined 
a critical point close
enough to the vertical axis of the phase diagram
that the dynamics of critical fluctuations are important
for a droplet of fluid with $\mu_B=0$ that expands and cools
following a trajectory down the vertical axis.  
(In reality, if there is a critical point in the
phase diagram of QCD it is at sufficiently large $\mu_B$ that
the dynamics of critical fluctuations are important
only in heavy ion collisions that produce droplets of 
fluid with sufficient, nonzero, $\mu_B$.)
  }
 \end{figure}
 %
 %

Since we are not aiming for a description of BES data in this paper, what is our goal?  We want to ``exercise Hydro+''; we want to see Hydro+ in action.  Our goal is to analyze the interplay between critical fluctuations and hydrodynamics in the simplest possible
model that we can set up where the dynamical feedback between the two
can be driven, with hydrodynamic expansion and cooling preventing the critical fluctuations from staying in equilibrium, and with
the out-of-equilibrium fluctuations driving the bulk dynamics
itself out-of-equilibrium also, and with each feeding back
upon the other.  
There is enough complexity in this goal that in many other
ways we shall make brutal simplifications.
As already noted, we will (i) not attempt to discern or employ
realistic initial conditions coming from the early stage dynamics
of a heavy ion collision at BES energies.  And, also as already
noted, we will (ii) make no
attempt to describe freezeout and particlization, and hence will 
make no predictions for experimental observables.
Furthermore, instead of doing a 3+1-dimensional hydrodynamic calculation we shall (iii) assume boost-invariant longitudinal
expansion and we shall (iv) assume that the expansion in the
transverse directions is azimuthally symmetric.  That is,
we analyze radial expansion dynamics (in the presence of
boost invariant longitudinal expansion) with no 
anisotropies.
This means that we must choose initial conditions that
are boost invariant and azimuthally symmetric, neither of 
which is a good representation of what will come from
the early stages of a heavy ion collision at BES energies.  
The final brutal simplification that we make is
that we shall (v) use an equilibrium EoS 
in which we place an imagined critical point 
near the $\mu_{B}=0$ axis of the phase diagram, as Fig.~\ref{fig:critical_point} illustrates. 
This allows us to do our entire calculation with $\mu_B=0$,
a considerable technical simplification.
In recent years, very substantial progress has
been made in relaxing all the simplifying assumptions (i)$\ldots$(v) in calculations done without critical fluctuations.
We anticipate that in future work it will be possible 
to meld these advances into our own.  But that is for the future.

We trust that it is apparent that our goal is not to do phenomenology. We shall provide a demonstration of how Hydro+ works in 
a setting in which all the physics that is unique to Hydro+ 
is manifest, and in an environment that is analogous to the
experimentally relevant setting modulo all the simplifications.
In addition to seeing how Hydro+ works we do expect
that, at a qualitative level, our results can provide 
some guidance for what to expect from future more complete simulations.
Features that we see in our results that we expect will generalize include: (a) out-of-equilibrium fluctuations lagging behind what they would be if they were able to stay in equilibrium; this qualitative feature
has been expected since Ref.~\cite{Berdnikov:1999ph} and indeed
we now see it manifest in this self-consistent setting; (b) nontrivial spatial dependence of the critical fluctuations originating 
because different regions of the droplet of hot fluid cool near the critical point at different times and subsequently shaped by
both critical slowing down and memory effects;
(c) advection of the critical fluctuations, with the radial flow in the fluid carrying them outwards; (d) imprints of the
out-of-equilibrium fluctuations on the hydrodynamic variables, including the energy density and radial flow.
Within our model, we find that the feedback on the hydrodynamic variables does not have dramatically large effects. 
Further work will be needed in order to check this quantitative conclusion in other settings but, if it persists, this
will considerably simplify future modelling.

This paper is organized as follows.
In Sec.~\ref{sec:hydro-plus-review},
we review the ingredients of Hydro+ which 
are pertinent to the present study~\cite{Stephanov:2017ghc},
and elaborate on various subtleties and practicalities that must be faced in any explicit implementation of Hydro+ that have not been treated previously.
We specify our model setup in Sections~\ref{sec:model}, \ref{sec:crit}, \ref{sec:EoS} and \ref{sec:IC},
in particular describing how we implement Fig.~\ref{fig:critical_point} and relate it to the Hydro+ equations in Sections~\ref{sec:crit} and \ref{sec:EoS} and describing how we initialize our calculation in Section~\ref{sec:IC}.
A reader who is only interested in results, or familiar with the Hydro+ formalism, can jump to Section~\ref{sec:results}, 
where we show the results of our simulations of both the out-of-equilibrium critical fluctuations and the out-of-equilibrium bulk dynamics, in the latter case looking at the deviation of the entropy density, energy density and radial flow from their 
equilibrium values.
We conclude and look ahead in Section~\ref{sec:conclusion}.

\section{A review of Hydro+ and our model}
\label{sec:Section2}

\subsection{A brief review of Hydro+}
\label{sec:hydro-plus-review}

\begin{figure} 
\center
\includegraphics[width=0.5\textwidth]{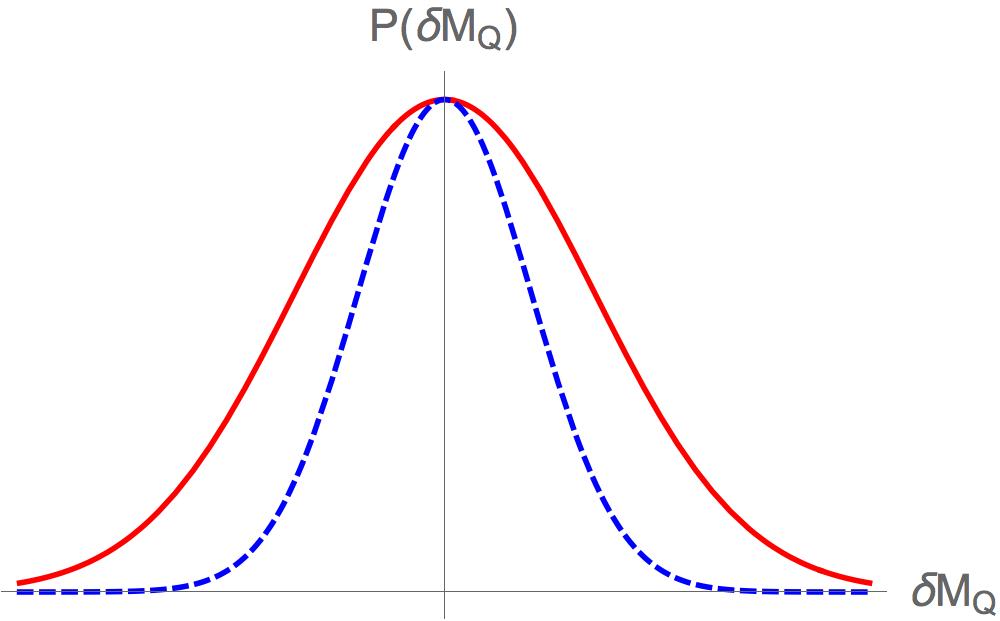}
\caption{
\label{fig:phiQ-plot}
A schematic illustration of the physical meaning of $\phi(Q)$. 
Here, we plot the probability distribution $P(\delta M_Q)$
of the fluctuation of the 
order parameter field $\delta M(Q)$ at momentum $Q$. 
$\phi(Q)$ corresponds to the width of the distribution $P(\delta M_Q)$.  That is, the red curve corresponds to a larger
$\phi(Q)$ than the blue dashed curve.
If we think ahead to future calculations, the order parameter $M$ 
which here at $\mu_B=0$ 
would just be the chiral order parameter $\langle \bar \psi \psi \rangle$ becomes a linear combination of the chiral order parameter
and the net baryon density.  The 
best way to understand how its fluctuations have observable consequences
is to note that the mass of the proton (and the neutron but neutrons are not seen in the experiments) is proportional to the chiral
condensate and hence $\delta M$ describes fluctuations in
both the net baryon density and the mass of the proton, each
of which correspond to fluctuations in the proton multiplicity
at freezeout.  
This picture, in quantitative form, can be used to estimate the
magnitudes of the contributions of fluctuations in the order parameter near a critical point
to observable fluctuations of quantities measured in experiments~\cite{Stephanov:1999zu,Hatta:2003wn},
in particular for the case of the
non-Gaussian cumulants of the event-by-event distribution of the
number of protons~\cite{Stephanov:2008qz,Athanasiou:2010kw,Stephanov:2011pb}.
That said, we remind the reader that we shall not
analyze freezeout in this paper.
}
\end{figure}

The primary goal of Hydro+ is to study the dynamics of critical fluctuations and their influence on the bulk evolution of a fluid near the critical point~\cite{Stephanov:2017ghc}.
Specifically, we consider the Wigner transform of the equal-time two point function of the fluctuation of an order parameter field $M(t,\vx)$:
\begin{eqnarray}
\label{phi-def}
\phi_{\vQ}\(t,\vx\)
\equiv \int d^3{\vy}\, \<\d M\(t,\vx-\vy/2\)\,\d M\(t,\vx+\vy/2\)\>\, e^{-i\vy\cdot\vQ}\, ,
\end{eqnarray}
where
\begin{eqnarray}
\label{dM-def}
\delta M\(t,\vx\)
\equiv M\(t,\vx\) -\<M\(t,\vx\)\>\, , 
\end{eqnarray}
with $\<\ldots\>$ denoting the ensemble average. 
$\phi_{\vQ}(t,\vx)$ describes the width of the probability distribution of $\d M$ at wavelength $1/Q$ for a subsystem of a fluid labelled by coordinate $\vx$ at given time $t$, 
see Fig.~\ref{fig:phiQ-plot} for an illustration. 
We note that $\phi_{\vQ}\(t,\vx\)$ in \eqref{phi-def} is defined in the local rest frame of the fluid.
The subtlety of defining the equal-time correlator in the presence of a nontrivial flow profile $u^{\mu}(t,\vx)$ was recently discussed in Ref.~\cite{An:2019osr}.

%
%
%

In Hydro+, $\phi_{\vQ}\(t,x\)$ is treated as a dynamical variable and obeys a relaxation rate equation\footnote{The definition of the relaxation rate $\Gamma_Q$ here differs from that in Ref.~\cite{Stephanov:2017ghc} by a factor of $\phi_{Q}/\phieq_{Q}$.}:
\begin{eqnarray}
\label{phi-eqn}
D\,\phi_{\vQ}(t,x)
= -
\Gamma_{\vQ}
\left( \phi_{\vQ}- \phieq_{Q}  \right)\,. 
\end{eqnarray}
In \eqref{phi-eqn}, we have defined\footnote{
According to Ref.~\cite{An:2019osr},  $D$ here has to be defined as the ``confluent derivative'', 
see Ref.~\cite{An:2019osr} for more discussion. 
We will not consider this refinement here.
}.  
\begin{eqnarray}
\label{D-def}
D\equiv u^\mu\, \pd_{\mu}\ . 
\end{eqnarray}
Therefore the evolution of $\phi_{\vQ}$ can be influenced by the effect of advection, 
as we shall illustrate through our simulation later. 
We will discuss the constraints on the behavior of $\phieq_{Q}$ and $\Gamma_{\vQ}$ coming from the critical universality in Section~\ref{sec:crit}. 

The conservation laws governing the dynamics of the standard hydrodynamic variables $\varepsilon$ (the energy density), $u^\mu$ (the fluid four-velocity) and $n_B$ (the baryon number density) take the usual form
\begin{eqnarray}
\label{conservation}
\pd_{\mu}T^{\mu\nu}=0\, ,
\\
\pd_{\mu}J^{\mu}=0\, ,
\end{eqnarray}
where $T^{\mu\nu}$ is the stress-energy tensor
and $J^\mu$ is the baryon number current, each of which is related to the hydrodynamic variables via a constitutive relation.
A central attribute of Hydro+ is that the out-of-equilibrium evolution of $\phi$ will back-react on the bulk evolution. 
The way that this is described in Hydro+ is that the standard constitutive relation satisfied by  the stress-energy tensor is modified and becomes:
\begin{eqnarray}
\label{Tmunu}
T^{\mu\nu}=
\varepsilon u^{\mu}u^{\nu}+ p_{\plus}\,\Delta^{\mu\nu} 
-\eta_{\plus}
\sigma^{\mu\nu}- \zeta_{\plus}\,\Delta^{\mu\nu}\theta
\end{eqnarray}
where we have defined:
\begin{eqnarray}
\Delta^{\mu\nu}&\equiv&\(g^{\mu\nu}+u^{\mu}u^{\nu}\)\, 
\\
\sigma^{\mu\nu}&\equiv& \nabla^\mu u^\nu + \nabla^{\nu} u^\mu -\frac{2}{3}\Delta^{\mu\nu}\theta\,, 
\end{eqnarray}
and
\begin{eqnarray}
\label{theta-def}
\theta\equiv\pd\cdot u\, , 
\qquad
\nabla^{\mu}
\equiv\Delta^{\mu\nu}\,\pd_{\nu}\, .
\end{eqnarray}
The constitutive relation \eqref{Tmunu} can be obtained from the standard hydrodynamic constitutive relation
by replacing the standard pressure $p$ with $p_{\plus}$ and by replacing the standard shear viscosity $\eta$ and bulk viscosity $\zeta$ with $\eta_{\plus}$ and $\zeta_{\plus}$, respectively. 
Here, the generalized pressure $p_{\plus}$ (see more below) depends on the hydrodynamic variables $\varepsilon$ and $n_B$ and on the additional Hydro+ variable $\phi_{\vQ}$, with the back-reaction of the critical fluctuations on the bulk hydrodynamics being described by the dependence of
these quantities on $\phi_{\vQ}$. 
As explained in Ref.~\cite{Stephanov:2017ghc}, $\eta_{\plus}$ and $\zeta_{\plus}$ are 
in general different from their counterparts in the
hydrodynamic limit, $\eta$ and $\zeta$. 
This is because the shear and bulk viscosity
receive additional contributions originating from the slow 
relaxation of $\phi$.   These additional contributions
vanish in equilibrium. Note also that 
the difference between $\eta$ and $\eta_{\plus}$ is 
suppressed because $\phi$ is a scalar function, but the difference 
between $\zeta$ and $\zeta_{\plus}$ is significant~\cite{Stephanov:2017ghc}. 
Since we shall be working exclusively at $\mu_B=0$, we will not need the constitutive relation for $J^{\mu}$, which can be found in Ref.~\cite{Stephanov:2017ghc}.

Remarkably, the functional dependence of $p_{\plus}$ on $\varepsilon$, $n_B$ and $\phi_{\vQ}$ can all be obtained explicitly~\cite{Stephanov:2017ghc},  
and this in turn allows for a self-consistent treatment of the back-reaction of $\phi$ on the hydrodynamic evolution. 
According to Ref.~\cite{Stephanov:2017ghc},
$p_{\plus}$ is related to the generalized entropy density $s_{\plus}$  by generalized thermodynamic relations,
c.f.~\eqref{p-plus} below. 
The generalized entropy density can be expressed as $s_{\plus}=s+\Delta s$, 
where $s$ is the ordinary entropy density and where $\Delta s$ depends on $\varepsilon$, $n_{B}$ and $\phi_{\vQ}$ according to~\cite{Stephanov:2017ghc}:
\begin{eqnarray}
\label{Delta-s}
\Delta\, s& =&
\frac{1}{2}\,\int_{\vQ}\, 
\[\log\(\frac{\phi_{\vQ}}{\phieq_{\vQ}}\)-\frac{\phi_{\vQ}}{\phieq_{\vQ}}+1\] 
\end{eqnarray}
where $\phieq$ is the equilibrium value of $\phi$, 
and where we have introduced the short-handed notation $\int_{\vQ}\equiv \int d^3Q/(2\pi)^3$.
We note that 
$\Delta s=0$ if $\phi_{\vQ}=\phieq_{\vQ}$ and
we also
note that $\Delta s$ is negative at any $\phi_{\vQ}$. 
These properties are necessary 
since the entropy has to reach its maximum possible value
when the system is in equilibrium. 
$s_{\plus}$ and $p_{\plus}$ are then related by
the generalized thermodynamic relation~\cite{Stephanov:2017ghc}
\begin{eqnarray}
\label{p-plus}
p_{\plus} = \beta^{-1}_{\plus}\, s_{\plus} - \varepsilon\, , 
\end{eqnarray}
where we have set $n_B=0$ and defined the generalized (inverse) temperature
\begin{eqnarray}
\label{eq:betaplus}
\beta_{\plus}
&\equiv&
\(\frac{\pd s_{\plus}}{\pd\varepsilon}\)_{\phi_Q}
=\b\(\varepsilon\)+\Delta\b\, ,
\end{eqnarray}
with
\begin{eqnarray}
\label{Delta-beta}
\Delta\b&=&
\frac{1}{2}\,\int_{\vQ}\, \frac{\pd\log\phieq_{Q}}{\pd\varepsilon}\,\(\frac{\phi_{\vQ}}{\phieq_Q}-1\)\, .
\end{eqnarray}
The generalized pressure $p_{\plus}$ appearing in \eqref{Tmunu} is then given by $p_{\plus}=p+\Delta p$ with:
\begin{eqnarray}
\label{Delta-p}
\Delta p = \frac{T\(\Delta s- w\Delta \b\)}{1+T\Delta\beta}\, ,
\end{eqnarray}
where the enthalpy is defined via $w\equiv\varepsilon+p$ as usual.
At a formal level, this completes the specification of Hydro+.
Our goal in this paper is to flesh this formalism out, turning
it into equations that we shall solve, in a model context that we
describe over the course of the next four subsections.

Before continuing, let us make one cautionary remark
about a sense in which our notation is misleading.  
The quantity $\Delta p$ that we have defined
describes the modification of the pressure relative to its
equilibrium value that is a consequence of
 the deviation of $\phi_{\vQ}$ away from
its equilibrium value, but as
we shall show in Appendix~\ref{sec:qualitative-discussion} 
it also describes a modification of the bulk viscosity arising from
the same dynamics.
The actual bulk viscosity, namely the term that multiplies
$\Delta^{\mu\nu}\theta$ in the stress-energy tensor \eqref{Tmunu} --- 
we shall refer to it
as the effective bulk viscosity $\zeta_{\rm eff}$ --- is
the sum of the $\zeta_{\plus}$ in $T^{\mu\nu}$ and 
the contribution to $\zeta$ coming from the out-of-equilibrium dynamics
of $\phi_{\vQ}$ that is hidden within $\Delta p$.
We illustrate this explicitly in Appendix~\ref{sec:qualitative-discussion}.
Note also that in our model calculation we shall choose initial conditions 
with
$\zeta_{\plus}=0$, which is maintained by the time evolution, 
meaning that the only source of bulk viscosity in our model
calculation will be that described in Appendix~\ref{sec:qualitative-discussion}.

\subsection{Our model and the Hydro+ equations we solve}

\label{sec:model}

Our goal in this paper is to set up a model that illustrates the nontrivial effects originating from the Hydro+ equations in a setting that at least resembles what would be needed to simulate heavy ion collisions, 
but to retain a level of simplicity that allows us to simplify our numerical calculations and interpret our results in such a way that we can clearly see Hydro+ in action.  In the Introduction, we have described the five simplifying assumptions that we shall make in order to achieve
these goals.
We shall place a critical point near $\mu_B=0$ as in Fig.~\ref{fig:critical_point} and only consider the dynamics of a cooling droplet of QGP with $\mu_B=0$, namely undoped QGP with zero net baryon number.  
This allows us to drop baryon density $n_{B}$ from the list of dynamic variables, and set the net baryon current $J^\mu$ to zero.
We therefore ``only'' need to solve 
\eqref{phi-eqn} and \eqref{conservation}, which we rewrite here in a more explicit form:
\begin{eqnarray}
\label{phi-eqn0}
D\,\phi_{\vQ}(t,x)
= -
\Gamma_{\vQ}
\left( \phi_{\vQ}- \phieq_{Q}  \right)\,. 
\end{eqnarray}
and
\bes
\label{hydro-eq-we-use}
\begin{eqnarray}
D\, \varepsilon &=& -\(\varepsilon+p_{\plus}\)\theta + \frac{1}{2}\Pi^{\mu\nu}\, \sigma_{\nu\mu}\, , 
\\ \label{acceleration-fluid}
\(\varepsilon+p_{\plus}\)\, D \, u^{\mu}
&=&\nabla^{\mu}p_{\plus} - \Delta^{\mu}_{\nu}\nabla_{\sigma}\Pi^{\nu\sigma}
+ \Pi^{\mu\nu}D\, u_{\nu}\, ,
\\
\label{Pi-eq}
\tau_{\Pi}\, \Delta^{\mu}_{\alpha}\,\Delta^{\nu}_{\beta}\, D\, \Pi^{\a\b}&=&
-\Pi^{\mu\nu}+\eta_{\plus}\,\sigma^{\mu\nu} - \tau_{\Pi}\, 
\( \Pi^{\alpha\mu}\,\omega^{\nu}_{\alpha}+\Pi^{\alpha\nu}\,\omega^{\mu}_{\alpha} \)\, . 
\end{eqnarray}
\ees
In \eqref{hydro-eq-we-use}
we have introduced one more tensor made from gradients of the fluid velocity $u^{\mu}$:
\begin{eqnarray}
\o_{\mu\nu}
&\equiv&
\frac{1}{2}\Delta^{\mu\a}\Delta^{\nu\b}
\( \nabla_{\b}u_{\a}-\nabla_{\a}u_{\b}\)\, .
\end{eqnarray}
And, in \eqref{hydro-eq-we-use} we have replaced $p$ and $\eta$ in the standard Muller-Israel-Stewart second order viscous hydrodynamic equation as considered in Ref.~\cite{Baier:2006gy} by $p_{\plus}$ and $\eta_{\plus}$. Again following the Muller-Israel-Stewart
formalism, we have introduced
the shear tensor $\Pi^{\mu\nu}$ which satisfies $u_{\mu}\Pi^{\mu\nu}=0$ and $\Pi^{\mu}_{\mu}=0$ and obeys a relaxation equation~\eqref{Pi-eq} that we shall discuss momentarily. 
Following Ref.~\cite{Baier:2006gy}, we have introduced
the second order transport coefficients $\tau_\Pi$ 
which is referred to as the shear relaxation time.
In the limit that $\tau_{\Pi}\to 0$, we see from \eqref{Pi-eq} that
$\Pi^{\mu\nu}\rightarrow \eta_{\plus}\sigma^{\mu\nu}$ which
means that
$T^{\mu\nu}$ will approach that given by \eqref{Tmunu} (with zero
bulk viscosity) 
and the hydrodynamic equations \eqref{hydro-eq-we-use} that
we use reduce to the equations of first order viscous hydrodynamics. 
However, as has been understood since the work of Israel and Stewart, even though we are not interested in physics to second order in gradients we must keep $\tau_{\Pi}$ finite in order to ensure causality in the numerical evolution of the hydrodynamic equations. 
We shall use $T\, \tau_{\Pi}= 4\eta_{\plus}/s$. Note that for a massless Boltzmann gas at vanishing coupling, one calculates $\tau_{\Pi}T= 6\eta/s$ \cite{Baier:2006um}, whereas infinite-coupling calculations in $\mathcal{N}=4$ SYM at large $N_c$ give $\tau_{\Pi}T=(4-2\ln 2)\eta/s=2.63\eta/s$ \cite{Baier:2007ix}.
Treating these as respective weak- and strong-coupling estimates for $\tau_\Pi$, we have chosen our value to lie between these two. Note that the value of $\tau_{\Pi}$
will have little effect on the dynamics; its role is
to serve as a causality-preserving regulator for
M\"uller-Israel-Stewart second-order viscous hydrodynamics. 

Unfortunately, we encounter a numerical instability in our solution of \eqref{hydro-eq-we-use}. When we write \eqref{Pi-eq} for an azimuthally symmetric, boost-invariant system, we find two terms proportional to $1/r$, $\Pi^\eta_\eta/r$ and $(2-v_r^2)\Pi^r_r/r$ (see Eq.~(6) of Ref.~\cite{Baier:2006um}). As $r \to 0$ in our simulation, we find that these two terms are a source of numerical noise, so we set them to zero when $r<.12$fm. We find that, other than fixing the instability, this procedure leaves our results unaffected.

As we explained earlier in Sec.~\ref{sec:hydro-plus-review}, 
the difference between $\eta$ and $\eta_{\plus}$ is suppressed because $\phi$ is a scalar function. 
We will take
\begin{eqnarray}
\frac{\eta_{\plus}}{s}= \frac{\eta}{s} = \frac{1}{4\pi}\, , 
\end{eqnarray}
a reasonable value within the range motivated by comparisons between
experimental measurement of, and hydrodynamic simulation of, anisotropic flow; see for example Refs.~\cite{Bernhard:2016tnd,Romatschke:2017ejr}. 
While we have set $\zeta_{\plus}=0$, the relaxation of $\phi_{\vQ}$ still leads to an effective bulk viscosity, see \eqref{zeta-eff} as discussed in Appendix.~\ref{sec:qualitative-discussion}.

As already noted in the Introduction, both to reduce computational cost and because doing so does not compromise any of 
our goals in this paper we 
consider a droplet of QGP whose longitudinal expansion is boost invariant and whose radial expansion transverse to the beam direction is azimuthally symmetric. Therefore local variables in our model will not depend on the spacetime rapidity $\eta_{s}$ or the azimuthal angle $\phi$, but will depend on the proper time $\tau$ and the radial coordinate in transverse plane $r$. 
We solve the Hydro+ equations numerically using a method very similar to that explained in Ref.~\cite{Baier:2006gy}\footnote{We have developed our Hydro+ codes based on the VH1+1 hydrodynamic code which was written and made public by Paul Romatschke \cite{Baier:2006um, Baier:2006gy, Romatschke:2007jx}}.

In the Sections that follow, we will specify inputs needed to solve \eqref{phi-eqn0} and \eqref{hydro-eq-we-use}.
To study the critical phenomenon, 
we wish to place a hypothetical critical point near the $T$-axis so that critical fluctuations may grow there, i.e., $\phieq$ and $\xi$ would grow around the critical temperature $T_c$.
We implement this in Section~\ref{sec:crit}, 
where we continue the specification of our model by describing how we introduce and parameterize $\phieq(Q)$, which describes the critical fluctuations as they would be if they were in equilibrium, as well as the equilibration rate $\Gamma(Q)$, both of which are needed in \eqref{phi-eqn0}. Both depend on the equilibrium correlation length $\xi$, meaning that we will need to specify our model for how $\xi$ depends on temperature.
We complete the specification of our model in Sections~\ref{sec:EoS} and \ref{sec:IC}.
In Section~\ref{sec:EoS}, 
we introduce the EoS that we use that incorporates the presence of a critical point near $\mu_B=0$ as in Fig.~\ref{fig:critical_point}, 
and in particular
discuss how the growth of the correlation length $\xi$ would 
influence the behavior of the EoS near $T_c$.
In Section~\ref{sec:IC},
we choose the initial conditions that we shall employ in our 
Hydro+ calculations.
In each of these Sections, we shall make further simplifying assumptions, some of which we have already mentioned in the Introduction. At all stages we will recall that we are trying to
achieve a setting in which we can see Hydro+ in action, watch the back-reaction between the critical fluctuations and the hydrodynamic variables that Hydro+ is designed to describe develop and ramify, and understand and assess their qualitative features.  The many simplifications that we employ in order to achieve these goals mean that our results cannot be compared to data. The methods that we develop, and the insights that we gain, will become key elements of future larger simulations with fewer simplifying assumptions, and    which incorporate treatments of the initial stages of the collision and of freezeout.

\subsection{The parameterization of $\phieq(Q)$, $\Gamma(Q)$ and the correlation length $\xi$
\label{sec:crit}
}

The equilibrium value of $\phi_{\vQ}$, 
$\phieq_{Q}$, can be written as
\begin{eqnarray}
\label{phi-equil}
\phieq_{Q}
&=& \chi_M\,
f_{2}\(Q\xi\)\, , 
\end{eqnarray}
where the susceptibility of the order parameter field in the zero momentum limit $\chi_M$ scales with $\xi$ as
\begin{eqnarray}
\chi_M\sim \xi^{2-\widetilde{\eta}}
\end{eqnarray}
with $\widetilde{\eta}$ being the corresponding critical exponent\footnote{The critical exponent that we denote $\widetilde{\eta}$ is conventionally denoted by $\eta$ in literature on critical phenomena. In this paper, we use $\eta$ to denote the shear viscosity.}.
(See, for example the textbook~\cite{onuki2002phase}.)
Here, $f_{2}(a)$ is an equilibrium universal scaling function~\cite{onuki2002phase} that we can choose to have unit normalization $f_2(0)=1$. 
$f_2(a)$ takes the asymptotic form~\cite{onuki2002phase} 
\begin{eqnarray}
f_2(a) \sim a^{-(2-\widetilde{\eta})}\, , 
\qquad a\gg 1\, .
\end{eqnarray}
From these universal considerations, we know the
behavior of $\phieq_{Q}$ at both small and large values of $Q$:
\begin{eqnarray}\label{eq:asymptoticsforphieq}
\phieq_{Q}&\sim &
\begin{cases}
\xi^{2-\widetilde{\eta}}\, ,
\qquad\qquad\qquad Q\leq \xi^{-1}\, ; 
\\
Q^{-(2-\widetilde{\eta})}\, , \qquad Q\gg \xi^{-1}\, . 
\end{cases}
\end{eqnarray}
Next, for simplicity we set the critical exponent $\widetilde{\eta}$ to zero, since its numerical value in the 3D Ising model, $\widetilde{\eta}\approx 0.036$~\cite{onuki2002phase},  is so small. That is, we simply parametrize $\chi_{M}$ as~\cite{onuki2002phase}
\begin{eqnarray}
\label{Chi-M}
\chi_{M}=c_{M}\, \xi^{2}\, . 
\end{eqnarray}
While $c_M$ will in general depend on temperature, we treat it as a constant because we assume that it does not vary much in the narrow band of temperatures around $T_c$ that are relevant to our considerations because at those temperatures $\xi$ is enhanced relative to its typical microscopic value, which we shall denote by $\xi_{0}$ and which we also take to be temperature-independent.
For the same reason, we will also treat $\G_{0}$ defined in \eqref{Gamma-xi-para} below as a constant. 

Our results for $\phi$ will all be proportional to $c_M$, but
we shall see later that the contribution of $\phi$ to the Hydro+ entropy density and other thermodynamic quantities is independent
of the value of $c_M$.
Also for simplicity, we will use the Ornstein-Zernike (OZ) form for $f_2$~\cite{onuki2002phase}
\begin{eqnarray}
\label{OZ}
f_2(a)=\frac{1}{1+a^2}\, .
\end{eqnarray}
which has the proper asymptotic behavior \eqref{eq:asymptoticsforphieq}. Consequently, the expression that we shall use for $\phieq_{Q}$ in our calculation is
\begin{eqnarray}
\label{phieq-model}
\phieq_{Q}= \frac{c_{M}\xi^{2}}{1+\(Q\xi\)^2}
=\frac{c_M}{Q^2+\xi^{-2}}\, . 
\end{eqnarray}
See Ref.~\cite{Guida:1996ep} for a more refined description of $\phieq$. In order to apply this expression in an explicit calculation, we shall need a model for how the correlation length $\xi$ depends on temperature.

Before turning to the correlation length itself,
we must specify how
the equilibration rate $\G_{Q}$ is related to $\xi$.  In general, $\G_{Q}$ will depend on $\varepsilon$ and $\phi_{\vQ}$.
When $\phi_{\vQ}$ approaches its equilibrium value, so does $\G_{\vQ}$:
\begin{eqnarray}
\label{Gamma-limit}
\lim_{\phi_Q\to \phieq_Q}\G_{Q}\to 
\Gammaeq_Q\, .
\end{eqnarray}
Here $\Gammaeq_{Q}$ will only depend on $\varepsilon$ and $Q$, 
and near the critical point it will depend on $Q$ only via
the combination $Q\xi$ meaning that it
can be parameterized as
\begin{eqnarray}
\label{Gammaeq-Q}
\Gammaeq_{Q}
= \Gamma_{\xi}\(\varepsilon\)\, f_{\G}(Q\xi)\, , 
\end{eqnarray}
Here, the characteristic relaxation $\G_{\xi}$ scales as
\begin{eqnarray}
\Gamma_{\xi}\sim \xi^{-z}\, , 
\end{eqnarray}
with $z>0$ being the dynamical critical exponent. 
In other words, 
the 
equilibration rate for modes with $Q\sim \xi^{-1}$ 
will vanish as $\xi$ approaches infinity.
This is the phenomenon of critical slowing down, and is the reason why it is impossible for critical fluctuations to stay in equilibrium arbitrarily near a critical point unless the system spends an arbitrarily long time there, which is certainly not the 
case in heavy ion collisions. 
The dynamical universal function $f_{\G}(a)$
takes the asymptotic form
\begin{eqnarray}
\label{f-Gamma-asy}
f_{\G}(a) \sim a^{z}\, 
\qquad {\rm when}~a\gg 1\ ,
\end{eqnarray}
is of order $1$ when $a\sim 1$ and, in the 
universality class that we shall employ (see below), goes to a constant for $a\to 0$.
Therefore we find that $\Gammaeq$ has the following behavior:
\begin{eqnarray}
\label{GammaQ-asym}
\Gammaeq_{Q}&\sim&
\begin{cases}
\xi^{-z}\, ,\qquad\qquad  Q\sim \xi^{-1}\quad {\rm and}\quad Q \ll \xi^{-1}\, , 
\\
Q^{z}\, ,\,\, \qquad\qquad Q\gg \xi^{-1}\, . 
\end{cases}
\end{eqnarray}

For simplicity, and in the absence of any
better motivated options, we shall take
$\Gamma_{\vQ}=\Gammaeq_{Q}$, which trivially satisfies  \eqref{Gamma-limit}.
Next, what value shall we choose for the dynamical 
critical exponent $z$ and what form shall we choose for the
dynamical universal scaling function $f_{\G}(a)$ appearing in $\Gammaeq_{Q}$ in \eqref{Gammaeq-Q}? Both depend on the dynamical universality class of the critical point.
The QCD critical point is in the dynamical universality class of
Model H~\cite{Son:2004iv,Fujii:2004za}, 
according to the classification of Halperin and Hohenberg~\cite{RevModPhys.49.435}, 
meaning that it has $z\approx 3$.\footnote{When the critical point lies out in the phase diagram of QCD at a substantial nonzero value of $\mu_B$, its order parameter is 
a linear combination of the chiral condensate and $n_B$.
It is the fact that the order parameter incorporates a chiral condensate
component that is most important to understanding
the observable consequences of its fluctuations.
It is the fact that it incorporates a component that
is a conserved density that controls the dynamics of its fluctuations.
Its equilibration is eventually determined by the diffusion of baryon density, and it is the fact 
that this is conserved 
together with the nonlinear nature of hydrodynamics
which are responsible for the dynamical critical exponent taking 
on a value $z \approx 3$~\cite{Son:2004iv}.}
However, for a critical point close to $\mu_B=0$,
the critical fluctuations do not involve fluctuations in $n_B$ and 
the critical order
parameter is almost purely the chiral condensate, which is
not a conserved density.  Since chiral symmetry is explicitly broken
in QCD, the order parameter for a hypothetical critical point
near $\mu_B=0$ may also include small components of energy and entropy
density, which we shall neglect. 
The principal simplifications that assuming a critical point near $\mu_B=0$ brings us are that we need not include a mean $n_B$ in our hydrodynamics and  that fluctuations in $n_B$ are not enhanced.
Making this assumption also means that
the critical point is to a good approximation
in the dynamical universality class of Halperin and Hohenberg's Model A and the appropriate dynamical
critical exponent is $z=2$~\cite{Berdnikov:1999ph}.
So, we shall use $z=2$ in our calculations, and use
the Model A form of $f_{\G}(a)$~\cite{RevModPhys.49.435}, meaning that we take
\begin{eqnarray}
\label{Gamma-xi-para}
\Gamma_{\xi}&=&\Gamma_{0}\, \(\frac{\xi}{\xi_{0}}\)^{-2}\, ,
\qquad
f_{\Gamma}(a) =1+a^2\, . 
\end{eqnarray}
Here $\G_{0}$ is a constant, representing 
the microscopic relaxation rate away from the critical point 
and our choice of $f_{\Gamma}(a)$ captures the desired asymptotics  \eqref{f-Gamma-asy} and \eqref{GammaQ-asym}.
(Note that the small $Q$ asymptotics in \eqref{GammaQ-asym} is
that of Model A, as appropriate in our model calculation.  This behavior
is different in Model H, where $f_\Gamma(a)\propto a^2$ for $a\to 0$
and $\Gamma_Q \sim Q^2\xi^{2-z}$ for $Q\ll \xi^{-1}$.)

As a consequence of all these considerations, the equation
of motion \eqref{phi-eqn} for $\phi_{\vQ}$ becomes
\begin{eqnarray}\label{phi-eqn-v2}
D\phi_{\vQ}
&=&- \Gamma_{0}\(\frac{\xi}{\xi_{0}}\)^{-2}\, \[1+\(Q\xi\)^2\]\[\phi_{\vQ}-\phieq_{Q}\]\, , 
\end{eqnarray}
where $\phieq_{Q}$ is given by \eqref{phieq-model}.
We will examine the dependence of our results on choices of the constant $\G_{0}$ in Section~\ref{sec:results}.

To close our discussion of  the equation of motion \eqref{phi-eqn-v2} for $\phi_{\vQ}$
and make it fully specified, we need to parameterize
the equilibrium correlation length $\xi$ as a function of $\varepsilon$ or, as we shall choose, $T$. 
Since we have placed a critical point near $\mu=0$ at some $T_c$, 
as the droplet of plasma cools past the temperature $T_c$ 
the equilibrium correlation length $\xi$ will first rise
will then peak at a large but finite value $\xi_{\max}$, and will then fall.
When $|T-T_c|$ is much larger than the width of the critical regime,
which we shall denote by $\Delta T$, the equilibrium correlation length
falls to some microscopic length $\xi_{0}$;
placing the critical point near $\mu_B=0$ means that for a
droplet that cools down the $\mu_B=0$ axis the equilibrium
length peaks at a $\xi_{\max}$ that is much larger than $\xi_0$.
We shall describe our choice for $\xi(T)$ in an equation 
momentarily, but
it may be helpful to look ahead to the top-left panel
of Fig.~\ref{fig:EoS} to see it plotted.  
We shall choose a simple ansatz for $\xi$ that approaches $\xi_0$ away
from $T_c$, that peaks at $\xi_{\max}$ and, motivated
by a mean theory result, that has $\xi^{-2} \propto |T-T_c|$
for small $|T-T_c|$ if the critical point is very close
to $\mu_B$ meaning that $\xi_{\max}\gg \xi_0$. We choose:
\begin{eqnarray}
\label{xi-para}
\(\frac{\xi}{\xi_0}\)^{-2}
=
\sqrt{\tanh^2\( 
\frac{T-T_c}{{\Delta T}}\)\(1-\(\frac{\xi_{\max}}{\xi_0}\)^{-4}\)
+\(\frac{\xi_{\max}}{\xi_0}\)^{-4}
}\, .
\end{eqnarray}
In our calculations, 
we shall take
\begin{eqnarray}\label{ximax-xi0}
\frac{\xi_{\max}}{\xi_0}
&=&3\, ,
\qquad
\xi_{0}=1~\fm\, ,
\end{eqnarray}
and
\begin{eqnarray}
\label{Tc-DeltaT-model}
T_{c}=0.160~\GeV\, , 
\qquad
\Delta T=0.2 T_{c}\, .
\end{eqnarray}
In the top-left panel of Fig.~\ref{fig:EoS}, we plot $\xi/\xi_0$ as a function of $T$, with these values of $T_c$ and $\Delta T$.  This expression completes our explicit specification of $\Gamma_{Q}$
and the equation of motion \eqref{phi-eqn-v2} for $\phi_{\vQ}$.
Our parameterization \eqref{xi-para} of $\xi(T)$ will
also play into our discussion of the equation of state.

\subsection{Construction of the Equation of State}
\label{sec:EoS}

In this Section,
we discuss the generalized EoS, $p_{\plus}=p(\varepsilon)+\Delta p$,
given by \eqref{p-plus} which, together with
the inverse temperature \eqref{eq:betaplus} and \eqref{Delta-beta}, yields the expression \eqref{Delta-p}.
The only thing that remains in order to turn \eqref{Delta-p} into
an explicit specification of $p_{\plus}$ is the explicit
specification of the equilibrium equation of state
$p(\varepsilon)$ in our model.

We shall provide $p(\varepsilon)$ by starting from a specific $c_{V}(T)$, 
and then determining $p(\varepsilon)$ via the standard thermodynamic relations
\begin{eqnarray}
\label{EoS-from-cV}
s(T)
&=&\int^{T}_{0}\,dT'\, \frac{c_V\(T'\)}{T'}\,
\\
\varepsilon(T)\label{varepsilon-from-cV}
&=&\int^{T}_{0}\,dT'\, c_V(T')\, 
\\
p &=& \frac{s}{\beta} - \varepsilon\ .\label{second-law}
\end{eqnarray}
From $c_V(T)$ we can also obtain 
the square of the sound velocity, as it is given by
\begin{eqnarray}
\label{cs-sovercV}
c^2_s&=& \frac{s}{c_{V}}\, . 
\end{eqnarray}
Finally, we shall also need the standard thermodynamic relation
\begin{eqnarray}
\label{dp-inbeta-alpha}
dp&=& \beta \[ - w\, d\b + n_B\, d(\mu_B/T)\]\, . 
\end{eqnarray}

To proceed, we need an ansatz for $c_V(T)$, both the contribution
associated with the critical point and the non-critical contribution.  
We start with the critical contribution, and begin
from the textbook mean field theory result
 (e.g.~see Ref.~\cite{kardar2007statistical}) that near
a critical point $C_V \propto \xi$.
(By ``mean field theory result'' 
we mean that we substitute the mean field theory value for the exponent $\nu=1/2$ into the hyperscaling relation 
$\a= d\nu -2$ to obtain $\a=-1/2$ for spatial dimension $d=3$, from which it follows that $C_{V} \propto \xi^{-\a/\nu}=\xi$. As described in Ref.~\cite{kardar2007statistical}, this corresponds to including the effects of Gaussian fluctuations, as in the derivation of
the Hydro+ formalism~\cite{Stephanov:2017ghc}.)
In our calculations, we 
shall use the explicit form
\begin{equation}\label{CV-ansatz}
c_V^{\rm crit}(T) = \frac{1}{2}\,\frac{1}{\xi_0^3}\,\frac{\xi(T)}{\xi_0}
\end{equation}
and use the parameterization \eqref{xi-para} for $\xi(T)$.
The powers of $\xi_0$ come from dimensional analysis, and although
we shall give an argument momentarily for our choice of the prefactor $1/2$ let us start by noting that we do not actually know the value
of this constant of order unity.
We can argue for our choice as follows.  In the Ising model, the analogue of $C_V$ (namely the heat capacity defined upon holding the extensive thermodynamic variable fixed) is $C_M$ and in the mean field theory
for the Ising model this is given by~\cite{kardar2007statistical}
\begin{eqnarray}
\label{CM}
C_{M,\Ising}&=& \frac{1}{16\pi}\,\frac{1}{\xi^{3}_{0}}\,\(\frac{\xi}{\xi_{0}}\)\, ,
\end{eqnarray}
where here $\xi(T)$ and $\xi_0$ are the Ising model correlation
length and its microscopic value away from the Ising model 
critical point.
Mapping an expression like this from the Ising model
onto an expression for $c_V^{\rm crit}(T)$ in
our model necessarily involves unknown nonuniversal factors of order unity,
but there is one contribution to this 
factor that we can estimate: because $C_V$ involves two derivatives of the free energy with
respect to the temperature we can expect that the prefactor
that is introduced via the mapping onto our model includes
a factor of $(T_c/\Delta T)^2$, which we have set to 25, see \eqref{Tc-DeltaT-model}. We have guessed a value of $1/2$ for the 
prefactor in  
the ansatz \eqref{CV-ansatz} because $25/16\pi \approx 1/2$. 
Surely in future it will be possible to much improve on
this, but we have made other more brutal simplifying assumptions
elsewhere so for our purposes in this paper the ansatz \eqref{CV-ansatz} will suffice.

We expect that the contribution of critical fluctuations to $c_{V}$ will only become important near $T_c$ and 
that $c_{V}$ will approach that without a critical point away from $T_{c}$. 
We therefore construct $c_{V}(T)$ as follows:
\begin{eqnarray}
\label{cv-full}
c_{V}(T)
&=&
\begin{cases}
c_{V}^{\noCP}(T)\, , \qquad T\leq T_{L}\\
c_{V}^{\crit}(T) +\sum_{n=0}\, c_{n}\, \(\frac{T-T_c}{\Delta T}\)^{n} \, , \qquad  T_{L}\leq T\leq T_{H}\, ,\\
c_{V}^{\noCP}(T)  
\, , \qquad T\geq T_{H}\, , 
\end{cases}
\end{eqnarray}
where we shall choose to take
\begin{eqnarray}
\( T_L,T_H\) =\(T_c-\Delta T, T_c +\Delta T\)\, ,
\end{eqnarray}
with $T_c$ and $\Delta T$ as in \eqref{Tc-DeltaT-model}.
The coefficients $c_{n}$ are chosen to satisfy matching conditions at the boundaries $T=T_{L,H}$.
Specifically, we will require $\(c_{V}/T^3\)$ and its first two derivatives to be continuous at $T=T_{L,H}$. 
(The number of derivatives 
to be matched at each boundary is a matter of choice.)
In order to satisfy these six constraints, we 
fit nonzero values of the six coefficients $c_{0,1,\ldots,5}$
\footnote{
With input parameters specified as we describe below, the fitted values of these parameters that we employ are given by
$\{c_0,c_1,c_2,c_3,c_4,c_5\}=\{26.80,7.29,0.38,-0.27,-0.12,0.01\}$.
}.
In this operational way, we obtain an ansatz for the
contribution to $c_V(T)$ near the critical point that comes
from all the degrees of freedom other than the critical 
order parameter.
To apply this matching procedure, we first need an ansatz
for $c_V^{\noCP}(T)$ that we shall use for $T<T_L$ and $T>T_H$.
We will use the following ansatz:
\begin{eqnarray}
\label{cv-no-CP}
\frac{c^{\noCP}_{V}}{T^{3}}
&=&
\[\(\frac{a_H+a_L}{2}\)+ \(\frac{a_H-a_L}{2}\)\, \tanh\(\frac{T-T_{\CO}}{\Delta T_{\CO}}\)\]\, . 
\end{eqnarray}
This means that in the high and low temperature limits, $c^{\noCP}_{V}/T^3$ (and hence $c_V(T)/T^3$) will approach temperature independent constants $a_H$ and $a_L$, respectively. 
The crossover from the low temperature regime to the 
high temperature regime happens around $T=T_{\CO}$ with 
the crossover width given by $\Delta T_{\CO}$. 
We shall choose
\begin{eqnarray}
\label{TCO}
T_{\CO}=T_c\ \ {\rm and} \ \ \Delta T_{\CO}=0.6\,T_c\, . 
\end{eqnarray}
Note that $T_{\CO}$ can in principle be different from $ T_{c}$ --- there is no reason why the critical point needs to sit {\it precisely} at the midpoint of the crossover --- but we use \eqref{TCO} for convenience. 
The width of the crossover $\Delta T_{\CO}$ in $c^{\noCP}_{V}$ has to be larger than $\Delta T$, and we have chosen it to be larger by a factor of three.
To complete our specification of $c^{\noCP}_{V}$, we choose 
\begin{eqnarray}
\label{aLH}
a_L=0.1\, a_{\QGP}\, , 
\qquad
a_H=0.8\, a_{\QGP} 
\end{eqnarray}
with $a_{\QGP}$ the value of $c_{V}/T^{3}$ for the non-interacting ideal gas QGP, namely
\begin{eqnarray}
\label{aQGP}
a_{\QGP}
= \frac{4\pi^2 (N^2_c-1)+21\pi^2 N_f}{15}\, ,
\end{eqnarray}
where $N_c=3$ and $N_f=3$ are the number of colors and flavors, respectively.
Our choice of $a_{H}$ is motivated by the lattice QCD calculations which show that $c_{T}/T^{3}$ of QGP approaches $a_{\QGP}$ from below 
very slowly in the high temperature limit and is around 80\% of this
value over a wide range of temperatures.
For real QCD nuclear matter, $c_{V}/T^{3}$ will vanish exponentially in low temperature limit, for temperatures much below the mass of the lightest hadron. 
For numerical simplicity it is easier to pick a small nonzero
value of $a_L$ as we have done, but none of the results that we shall 
focus on depend on this choice.

%
%
%
\begin{figure} 
\center
%
%
\includegraphics[width=0.45\textwidth]{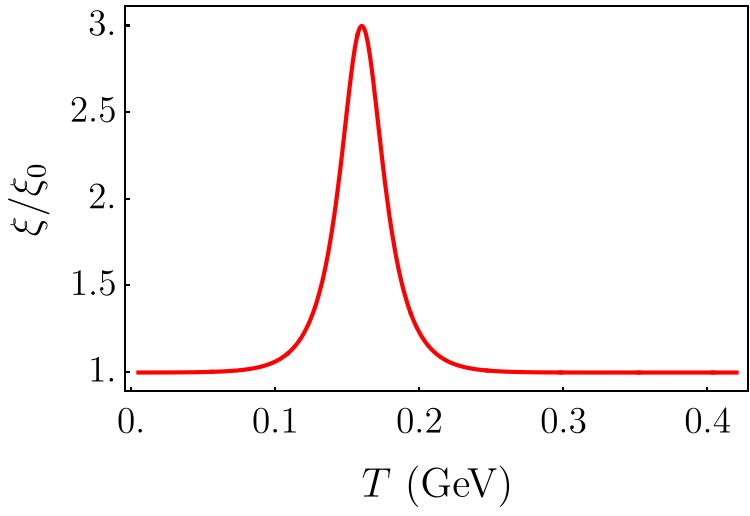}
\includegraphics[width=0.45\textwidth]{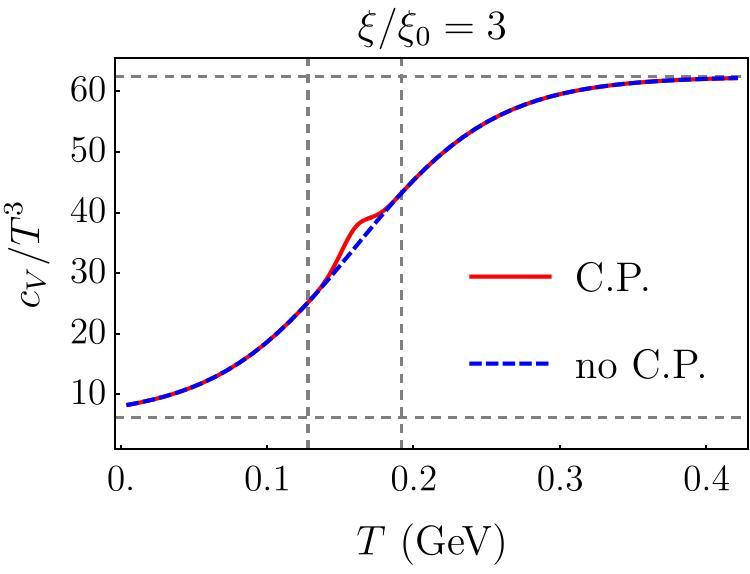}
%
%
\includegraphics[width=0.45\textwidth]{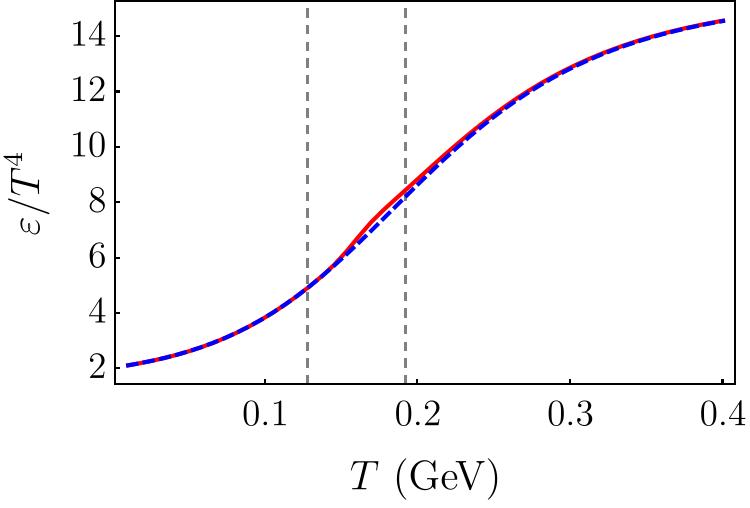}
\includegraphics[width=0.45\textwidth]{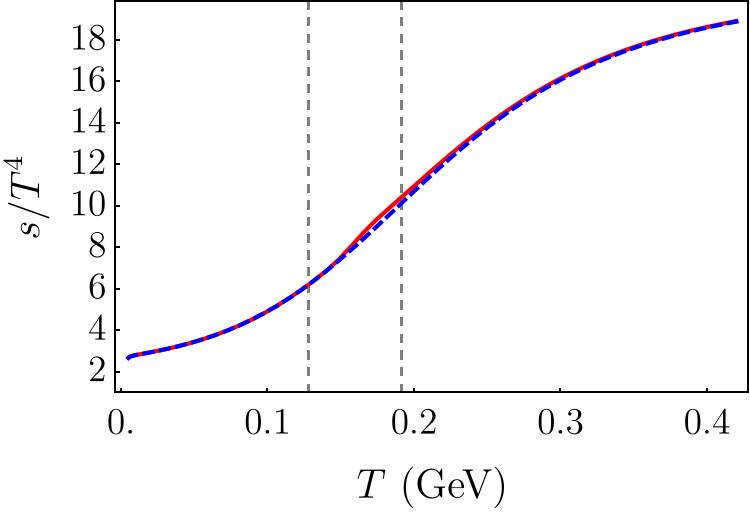}
%
%
\includegraphics[width=0.45\textwidth]{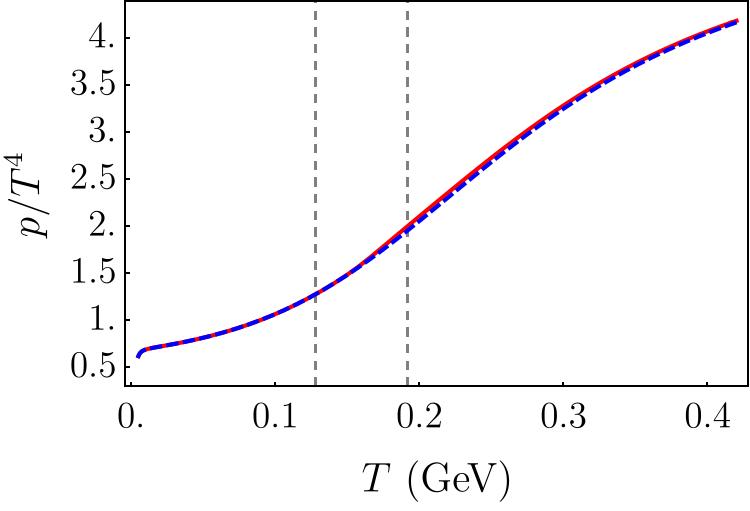}
\includegraphics[width=0.45\textwidth]{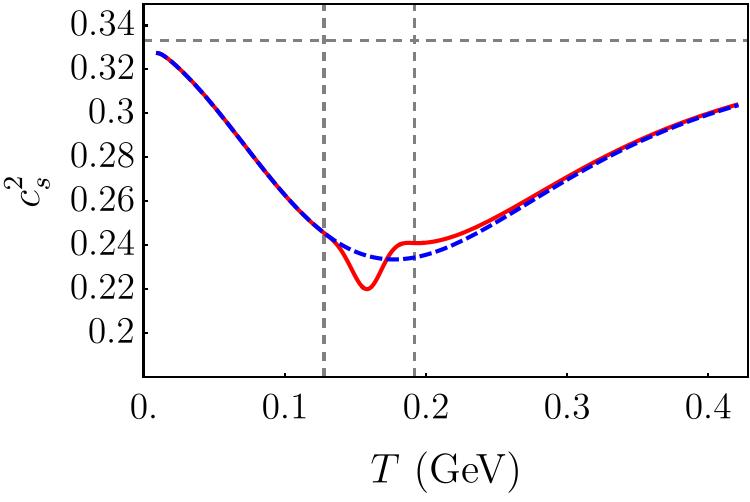}
\caption{
\label{fig:EoS}
The red curves in the upper two panels show the shapes of 
$\xi/\xi_0$ vs $T$ from \eqref{xi-para} and $c_{V}/T^3$ vs $T$
from \eqref{cv-full} that we have used to specify our model.
We have used $\xi_{\max}/\xi_{0}=3$.
The red curves in the middle and lower panels
show the consequent shapes of all the other thermodynamic
quantities:
$\varepsilon/T^4$ vs $T$, $s/T^3$ vs $T$,
$p/T^4$ vs $T$ and $c^{2}_{s}$ vs $T$.
In all the panels except for the top-left, the
blue dashed curves show the relevant thermodynamic
function without any critical contribution.
The left (right) vertical dashed line shows the location
of $T=T_L$ ($T= T_H$). 
}
\end{figure}
 %
%
%

We now pause to compare the magnitude of the critical contribution to $c_V(T)$ at $T=T_c$ to the non-critical contribution, as one way of checking
that all the ans\"atze we have made look reasonable.
From \eqref{CV-ansatz}, we have $c_V^{\rm crit}(T_c)= 1.5/\xi_0^3 = 1.5 \,{\rm fm}^{-3}$, where we have used the ansatz \eqref{ximax-xi0}, and hence
$c_V^{\rm crit}(T_c)/T_c^3 \approx 2.9$ for $T_c=160~{\rm MeV}\approx 0.8~{\rm fm}^{-1}$.
We can compare $c_V^{\rm crit}(T_c)/T_c^3 \approx 2.9$ to $a_{\QGP}\approx 62.5$, as follows.  $a_{\QGP}$ corresponds
to 16 bosonic degrees of freedom and 36 fermionic degrees of freedom, whereas $c_V^{\rm crit}(T_c)/T_c^3$ comes from a single scalar
order parameter degree of freedom whose contribution has been
enhanced at $T_c$ by a factor of $\xi_{\rm max}/\xi_0=3$.
If we take this comparison literally, it means that we have
slightly underestimated $c_V^{\rm crit}(T)$, just slightly.
Better to say that it gives us some confidence that the choices
we have made are not unreasonable.

We now have all the ingredients we need in order
to build our equation of state $p(\varepsilon)$ and all
the standard thermodynamic quantities.
We start from $c_V(T)$ given by \eqref{cv-full} and \eqref{CV-ansatz}
with $\xi(T)$ given by the ansatz \eqref{xi-para}
and then use \eqref{EoS-from-cV}, 
\eqref{varepsilon-from-cV}, \eqref{second-law} and \eqref{cs-sovercV}
to obtain $s$, $\varepsilon$, $p$ and the speed of sound $c_s$.
In Fig.~\ref{fig:EoS}, 
we plot $\xi(T)$ as well as $c_{V}(T)$, $s(T)$, $\varepsilon(T)$, $p(T)$, and $c^2_{s}(T)$ as functions of $T$ as the red solid curves. 
We also show all of the thermodynamic  quantities
without any critical contribution as the blue dashed curves.
Let us focus on $c^2_s$ which plays an important role in driving the hydrodynamic expansion. 
We observe as expected that $c^2_{s}$ vs $T$ 
features a minimum around $T=T_{c}$ since the equation of
state becomes soft near a critical point. 
We note that $c^{2}_{s}$ also shows a maximum around $T=T_{H}$.
To understand this, recall 
the relation $c^2_{s}=s/c_{V}$ (c.f.~\eqref{cs-sovercV}). 
$c_{V}$ has to decrease rapidly from its peak value 
to approach $c^{\noCP}_{V}$ around $T=T_{H}$, 
which in turns leads to a bump in $c^{2}_{s}$ vs $T$.

\subsection{Initial conditions}
\label{sec:IC}

We have now specified all the elements of our model that
we need in order to evolve the Hydro+ equations.  All that
remains in order for us to complete the full specification
of our model calculation is choosing 
the initial conditions for the time evolution.

We shall initialize our model at 
\begin{eqnarray}
\tau_{I}=1~\fm\, \,
\end{eqnarray}
with an initial central temperature of 330~MeV, following
Ref.~\cite{Baier:2006gy}. In future phenomenological modelling
of BES-energy heavy ion collisions, a somewhat lower initial temperature may be appropriate.  
Although we know these are not realistic assumptions, we shall follow
many authors including for example those of
Ref.~\cite{Baier:2006gy} in assuming that there is no radial flow and no $\Pi^{\mu\nu}$ initially, i.e. $v_{r}=\Pi^{\mu\nu}=0$ 
at $\tau=\tau_{I}$.
We will use the standard Glauber model corresponding to a central Au-Au central collision at $\sqrt{s}=200$~\GeV~for $\varepsilon(r)$ vs $r$
at $\tau=\tau_I$~\cite{Baier:2006gy}, and use the results from Section~\ref{sec:EoS}, see e.g.~Fig.~\ref{fig:EoS} to initialize the $r$-dependent values of all the other thermodynamic quantities.
For simplicity, we will assume that $\phi_{\vQ}$ is in equilibrium initially, i.e., $\phi_{\vQ}=\phieq_Q$. 
In at least one respect, this is likely unrealistic. At $\tau=\tau_I$ there will be some range of radii (at a relatively large $r$, near the edge of the fireball) where the QGP initially has a temperature near
$T_c$.  Our simplifying choice of initial conditions means that, for some range of $Q$, in this shell of radii we will have 
a large $\phi_{\vQ}$ from the start.  Although this is almost
certainly unrealistic, since there is no reason
to assume that $\phi_{\vQ}$ will have had time to reach its equilibrium value in this region, it will at the same time be 
very helpful in exercising the Hydro+ formalism, as we will be able
to watch how this feature in $\phi_{\vQ}$ evolves with time.
Note also that by virtue of our
choice of initial conditions $\phi$ in our simulation will depend initially only on the magnitude of $\vQ$, $Q$,
but not on its direction. And, as one can verify by inspecting \eqref{phi-eqn0}, this simplification will be maintained by the time
evolution.

We close this Section with a further remark about one
aspect of our initial conditions that can be improved in future.
Let us define $r_{L}$ and $r_{H}$ at each given $\tau$ through the conditions $\varepsilon\(r_{L}\)=\varepsilon\(T_{L}\)$ and $\varepsilon\(r_{H}\)=\varepsilon\(T_{H}\)$. 
That is, the QGP within the shell defined by
$r_{H}<r<r_{L}$ has a temperature that lies within
the range $T_L<T<T_H$.
It might be tempting to view the fluid in this shell
as critical, with a long correlation length, with $\xi=\xi_{\rm max}$
somewhere in the shell, as if it were in equilibrium. Indeed,
we have initialized $\phi_{\vQ}$ as if this were so at $\tau=\tau_I$. 
However, 
this could only make sense if $|r_{H}-r_{L}|> \xi_{\max}$,
and this condition is not satisfied at very early times (c.f.~Fig.~\ref{fig:bulk} below).
This means that a realistic initialization of $\phi_{\vQ}$ within this thin shell will require an analysis of finite size effects as well as
consideration of the shortness of $\tau_I$.
We leave this to future work, although we also note that
we expect that the observable consequences of the critical
fluctuations will be dominated by those throughout the interior of the
droplet which develop later, and whose development Hydro+ is
designed to describe.

\section{Results}
\label{sec:results}

The purpose of the present work is to demonstrate the intertwined dynamics among flow and critical fluctuations $\phi(Q)$. 
We shall present and describe results from our
model calculations of the evolution of $\phi(Q)$ in Sec.~\ref{sec:phi-ev}, and then 
in Sec.~\ref{sec:bulk-results}
we shall focus on the feedback of $\phi(Q)$ on the bulk 
hydrodynamic evolution.

%
%
%
\begin{figure}
\center
%
%
\includegraphics[width=0.465\textwidth]{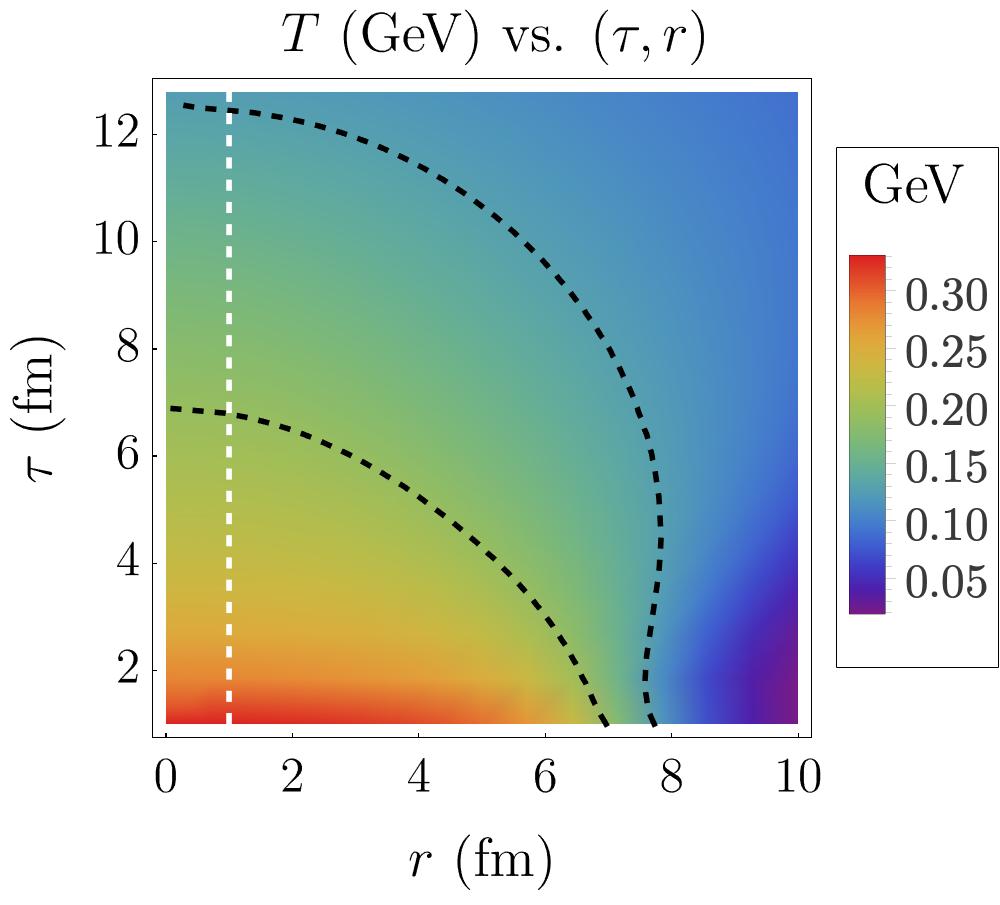}
\includegraphics[width=0.45\textwidth]{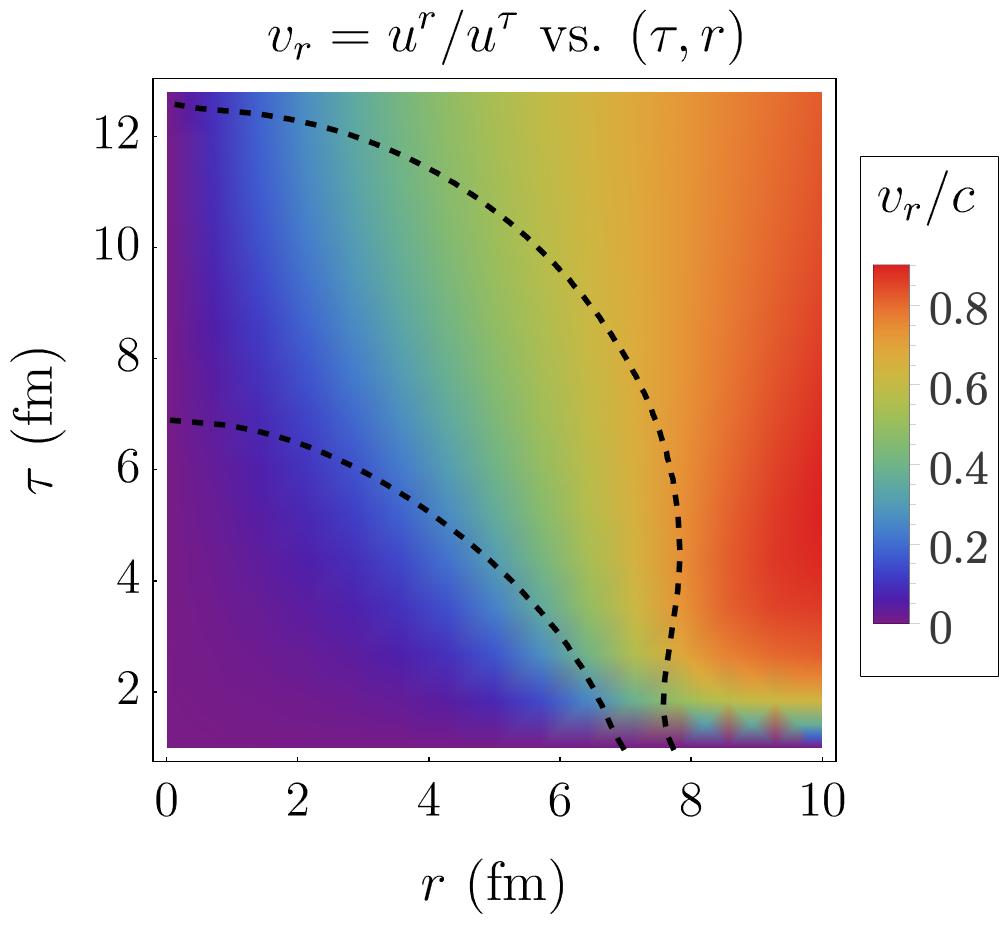}
%
%
\caption{
\label{fig:bulk}
The temperature $T$ and radial flow $v_r$ as functions of
the spacetime coordinates $(r,\tau)$ obtained from solving
standard hydrodynamic equations using the non-critical
EoS that we employ in our model.
We shall see as we go farther that including
a critical point and solving the Hydro+ equations 
changes the bulk evolution, but not dramatically.
Hence, these Figures are a good reference 
from which to get an initial sense of what our
model calculation describes.
In the Figure, 
the black dashed curves correspond, from below to above, to $T=T_H$ and $T=T_L$, respectively, where $T_H=1.2 T_c$ and $T_L=0.8 T_c$ with $T_c=160$~MeV.
They together bracket the 
critical regime where critical slowing down and Hydro+ dynamics 
should be expected to become important.
The vertical white dotted line corresponds to $r=1~\fm$,
a representative value of the radius at which we will illustrate the temporal evolution of $\phi(Q)$, 
see Figs.~\ref{fig:phi-fix-r} and~\ref{fig:phi-fix-r-diff-Gamma} below.
}
\end{figure}

As a preamble to the presentation of our results, however, we 
begin in Fig.~\ref{fig:bulk} by showing results for the 
spacetime evolution of the
temperature $T(\tau,r)$ and radial flow $v_{r}(\tau,r)$ 
obtained by solving standard
hydrodynamic equations 
using our non-critical model equation of state that we have
obtained in Section~\ref{sec:EoS} by starting from the
model \eqref{cv-no-CP} for $c_V(T)$ and applying standard
thermodynamic relations.
Solutions for these bulk variables obtained by using
the critical EoS from Section~\ref{sec:EoS} are similar because of the relatively small difference between the two equations of state, see Fig.~\ref{fig:EoS}. 
In Fig.~\ref{fig:bulk}, we observe familiar behavior of the temporal and spatial dependence of $T$ and $v_{r}$ for a fireball undergoing boost-invariant longitudinal expansion as well as radial flow.
At each $\tau$, the QGP fluid is hotter at smaller $r$. 
As the system expands and cools, the temperature (and energy density) drops while at the same time, the radial flow starts building up due to pressure gradients. 
To illustrate the boundary of the critical regime where critical slowing down is expected and the use of Hydro+ is necessary
we have shown the contours at which $T=T_H$ and $T=T_L$ as
black dashed curves in
 Fig.~\ref{fig:bulk}.

\subsection{The evolution of $\phi$}
\label{sec:phi-ev}

 %
%
%
%
%
%
\begin{figure}
\center
%
%
\includegraphics[width=0.45\textwidth]{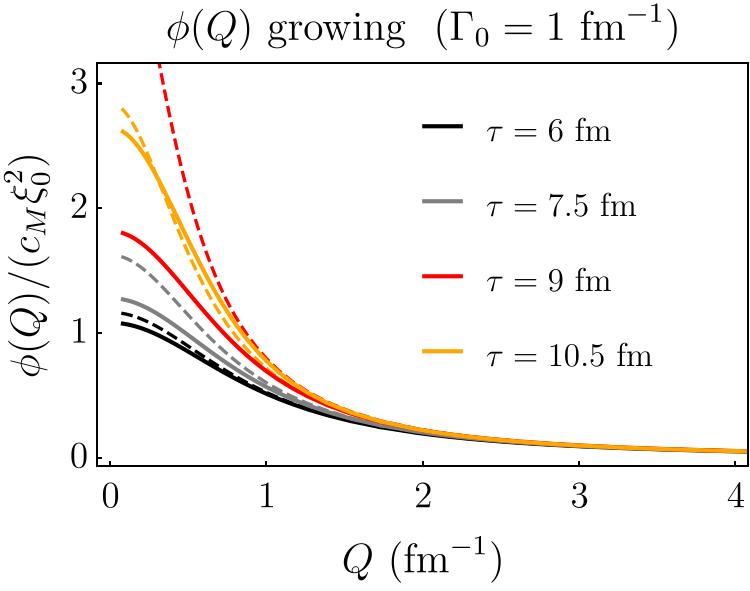}
\includegraphics[width=0.45\textwidth]{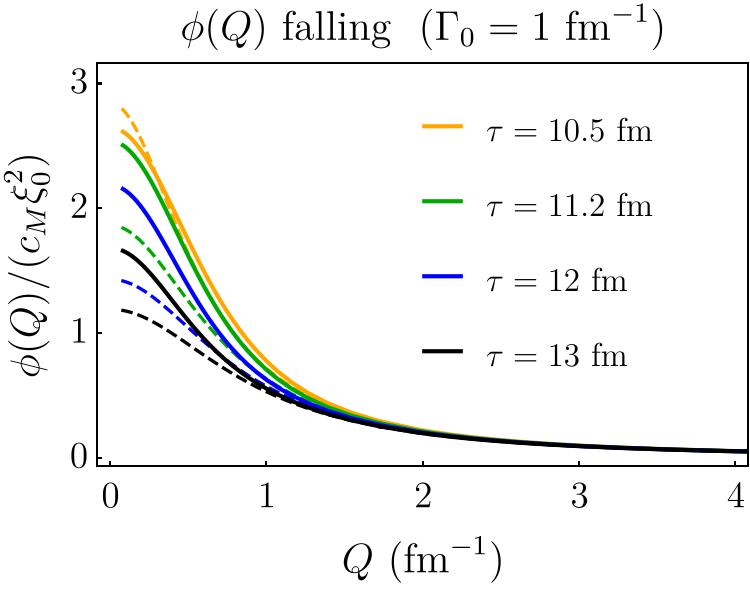}
%
%
%
\caption{
\label{fig:phi-fix-r}
The magnitude of the fluctuations of the critical order parameter with wave vector $Q$, $\phi(Q)$ defined in \eqref{phi-def}, plotted as a function of $Q$ at a representative radius $r=1$~fm.
We have calculated the dynamics of $\phi(Q)$
by solving the Hydro+ equations with $\G_{0}=1$~$\fm^{-1}$.
As the system cools through the critical regime, 
the equilibrium fluctuations ($\phieq(Q)$, shown as dashed curves)
first rise, and then fall.
The Hydro+ dynamics describe how $\phi(Q)$ (shown as solid curves) responds.
The ordinary hydrodynamics of an expanding cooling droplet
of plasma, in the presence of a critical point,
drives the dashed curves first upwards and then downwards.
As the dashed curves rise, the solid curves rise also, but
lag behind (first three values of $\tau$, shown in the left panel). At $\tau=10.5~\fm$ (both panels) the solid curve catches up to the dashed curve, though at low (high) $Q$ the solid curve lags below (above) because lower Q modes have smaller relaxation rates. As the dashed curve drops
further, the solid curves drop also, but lag behind (last three
values of $\tau$, shown in the right panel).
The lag, captured by the Hydro+ dynamics and seen in both panels, is a direct manifestation of critical slowing down.
}
\end{figure}
 %
%
%

\begin{figure}
\center
%

\begin{subfigure}
\centering
\includegraphics[width=0.45\textwidth]{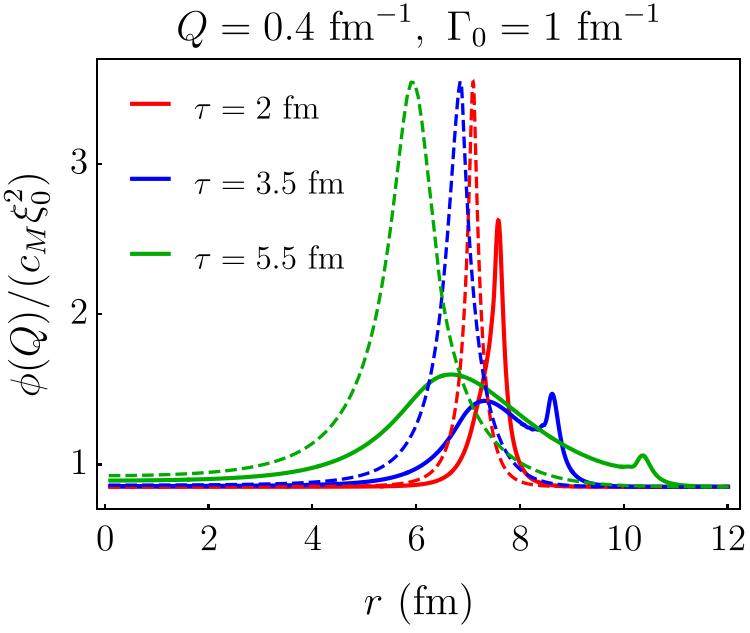}
\includegraphics[width=0.45\textwidth]{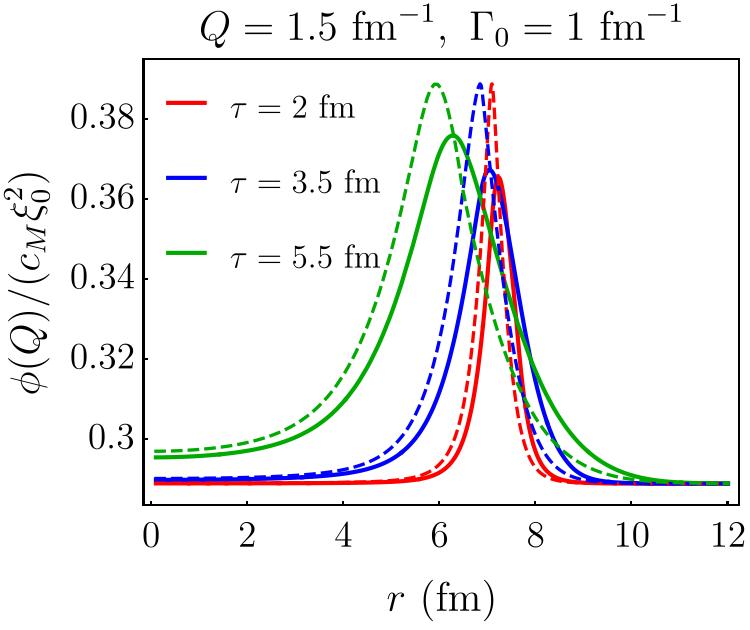}
\end{subfigure}

\begin{subfigure}
\centering
\includegraphics[width=0.45\textwidth]{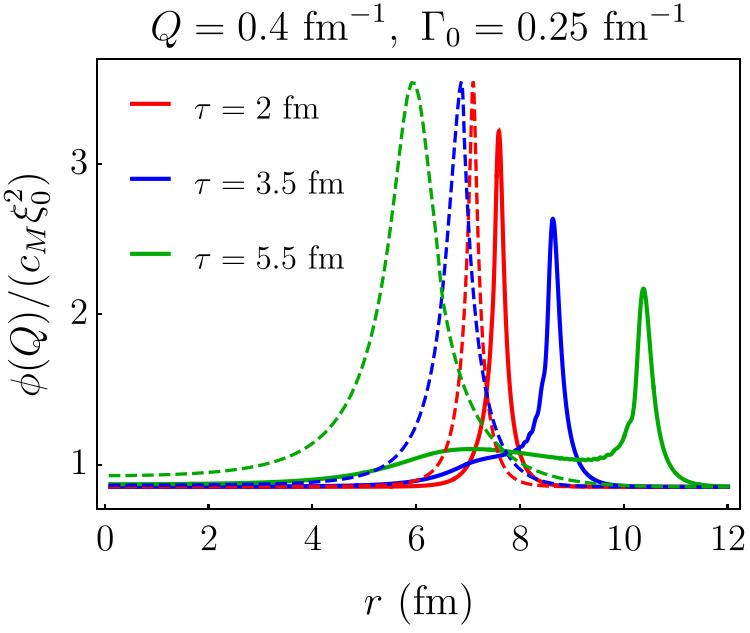}
\includegraphics[width=0.45\textwidth]{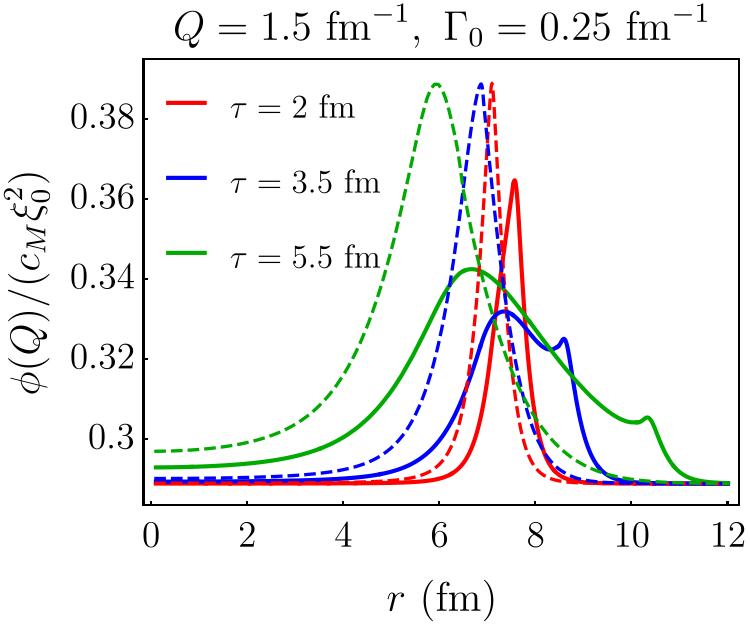}
\end{subfigure}
%
%
\caption{
\label{fig:phi-fix-Q}
The magnitude of the critical fluctuations, $\phi(Q)$, 
plotted as a function of radius $r$ at two values
of the wave vector, 
$Q=0.4~\fm^{-1}$ (two left panels) and $Q=1.5~\fm^{-1}$ (two right panels). Note that the vertical scales are different in the left and right panels, and note that zero is suppressed on all the vertical scales.
We have obtained these results 
by solving Hydro+ equations with two values of $\G_0$: $\G_{0}=1~\fm^{-1}$ (upper panels) and $\G_{0}=0.25~\fm^{-1}$ (lower panels).
In all panels, solid and dashed curves show $\phi(Q)$ and $\phieq(Q)$ respectively and 
the red, blue and green curves show results at $\tau=2$, $3.5$, and $5.5$~fm, respectively. 
Since at all times $\tau$ the fluid is hotter at smaller $r$, 
as time passes and the temperature of the fluid drops
the peak of $\phieq(Q)$ will move inward, toward smaller values of $r$.  At smaller values of $r$, therefore, we see $\phi(Q)$ rising as it tries to keep up with the rising $\phieq(Q)$.
In the case where $\G_0$ or $Q$ is sufficiently small, we see an additional effect:
the peak in $\phi(Q)$ present in the initial conditions that we have chosen in our model, namely the peak that is initially centered
at the values of $r$ that are in the critical regime,
is slow to relax and is carried outwards toward larger $r$
advectively by the radially flowing fluid.
}
\end{figure}

We begin the presentation of the results of our 
model calculation of $\phi(Q)$ by choosing a particular value of $r$ that is representative
of the interior of the fireball, $r=1~\fm$,
and plotting the temporal evolution of the critical fluctuations $\phi(Q)$ at this $r$ as a function of the momentum $Q$ in the critical regime.
In each of the two panels of Fig.~\ref{fig:phi-fix-r}, 
we choose four values of $\tau$  and plot $\phi(Q)$ at
each $\tau$, all at $r=1~\fm$. We have set
$\Gamma_{0}=1~\fm^{-1}$ in this calculation.
As the fireball cools, the temperature at $r=1~\fm$
(see the dashed white line in the left panel of Fig.~\ref{fig:bulk}) drops through the critical regime and the 
equilibrium value of $\phi(Q)$, namely $\phieq(Q)$ depicted
by the dashed curves in Fig.~\ref{fig:phi-fix-r}, first
rises as the temperature approaches $T_c$ from above (left
panel of Fig.~\ref{fig:phi-fix-r}), then reaches
a maximum value 
where $T$ passes $T_c$,
and finally falls as the temperature drops farther below $T_c$
(right panel of Fig.~\ref{fig:phi-fix-r}).
In the left panel of Fig.~\ref{fig:phi-fix-r}, we
see $\phi(Q)$, depicted by the solid curves,
rising but lagging behind $\phieq(Q)$, ``trying to catch up''.
At $\tau=10.5~\fm$ it catches up, because at this $\tau$ the equilibrium $\phieq(Q)$ has already turned around and started coming downward.
We show $\phi(Q)$ at $\tau=10.5~\fm$, when it is at its maximum
value, in both panels of Fig.~\ref{fig:phi-fix-Q}.
In the right panel of Fig.~\ref{fig:phi-fix-r}, we
see $\phi(Q)$ dropping, but again lagging behind $\phieq(Q)$ which now means that it is higher, again ``trying to catch up'' as $\phieq(Q)$ drops.
Note that what we have described as the solid curve lagging
behind the dashed curve can equally well be described
as a memory effect: the solid curve ``remembers'' 
where the dashed curve used to be, meaning that as the dashed curve rises the solid curve is below it and later when the dashed curve has dropped 
the solid curve ``remembers'' some of its former height.
Regardless of the pictorial language that one chooses,
the difference between the solid and dashed curves
is an illustration of the out-of-equilibrium physics
of the critical fluctuations that Hydro+ is designed
to describe and is a direct manifestation of critical
slowing down.

%
%
%
\begin{figure}
\center
%
%
\includegraphics[width=0.45\textwidth]{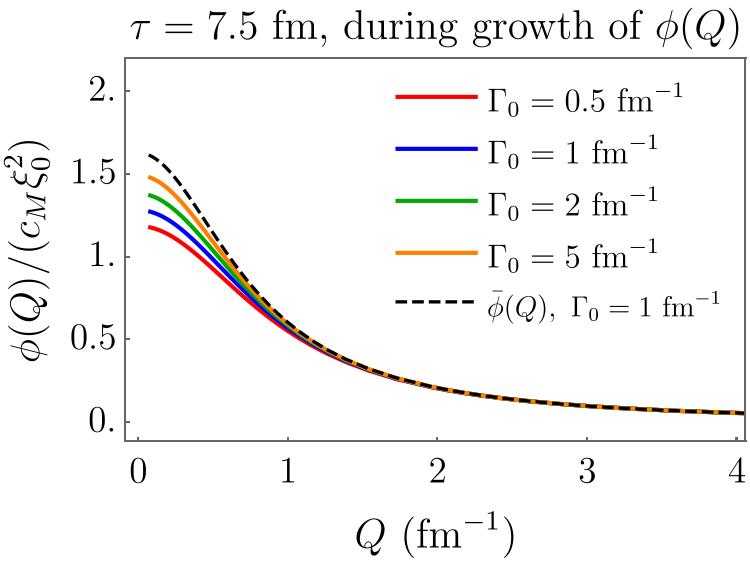}
\includegraphics[width=0.45\textwidth]{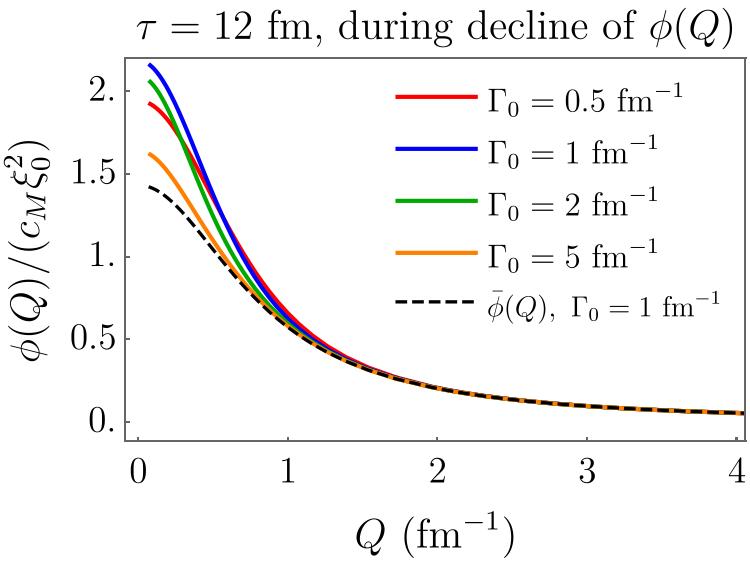}
\caption{
\label{fig:phi-fix-r-diff-Gamma}
The magnitude of the critical fluctuations, $\phi(Q)$, 
plotted as a function of their wave vector $Q$ 
at the same radius $r=1$~fm as in Fig.~\ref{fig:phi-fix-r}
at two values of $\tau$, $\tau=7.5~\fm$ (left panel) and
$\tau=12~\fm$ (right panel). In both panels, we show
results obtained by 
solving the Hydro+ equations for our model 
at four different values of $\G_{0}$: 
$\G_{0}=0.5, 1, 2, 5$~$\fm^{-1}$.
Solid and dashed curves show $\phi(Q)$ and $\phieq(Q)$ respectively.
In the left panel, the solid curves are rising, but lagging behind the dashed curve.  In the right panel, the solid curves are dropping, and again are lagging behind the dashed curve. In the left panel, the larger the value of $\G_0$ the less the lag, the closer the solid curve is to the dashed curve, and the smaller the out-of-equilibrium effects. In the right panel, the dependence of the out-of-equilibrium effects on the value of $\Gamma_0$ is more complex, as described in the text.}
\end{figure}
 %
%
%
%

To complement Fig.~\ref{fig:phi-fix-r}, 
in Fig.~\ref{fig:phi-fix-r-diff-Gamma}
we show $\phi(Q)$ at $r=1~\fm$ at two different values of $\tau$ obtained
from Hydro+ calculations done using four different values of $\G_{0}$.
In the left panel of Fig.~\ref{fig:phi-fix-r-diff-Gamma} we see that, as expected, the out-of-equilibrium effects
become smaller as $\G_0$ increases, since larger $\G_0$ means
more rapid relaxation toward equilibrium
and hence the larger the value of $\Gamma_0$ the more rapidly
the solid curve responds as the dashed curve moves, meaning
the less the solid curve lags behind the rising dashed curve.
In the right panel of Fig.~\ref{fig:phi-fix-r-diff-Gamma}, the dependence of the solid curves on the
value of $\Gamma_0$ arises from two effects: (i) the smaller the value of $\Gamma_0$ the more slowly the solid curve $\phi(Q)$ relaxes
down toward the dashed $\phieq(Q)$, the more the solid curve lags behind the dropping 
dashed curve, meaning the higher the solid curve is; (ii) for small
values of $\Gamma_0$ the solid curve never rose as far during
the earlier epoch when it was ``trying to follow'' the rising dashed curve,
meaning that during this epoch the solid curve is lower for smaller values of $\Gamma_0$. We see from the solid curves
in the right panel of Fig.~\ref{fig:phi-fix-r-diff-Gamma} that the first effect is dominant at larger $Q$ while the second effect is more significant at the lowest values of $Q$, meaning that the solid curves with different values of $\Gamma_0$ can cross each other as a function of $Q$.

We can also observe, in both Figs.~\ref{fig:phi-fix-r} and Fig.~\ref{fig:phi-fix-r-diff-Gamma}, 
that modes with a large enough wave vector $Q$ 
are always close to equilibrium, 
for any value of $\Gamma_{0}$ and $\tau$ under consideration. 
(The value of $Q$ that is ``large enough'' is smaller 
for larger values of $\Gamma_0$.)
In illustrating the out-of-equilibrium dynamics that Hydro+ describes,  therefore, we shall henceforth focus more on smaller values of $Q$, which is to say on the longer wavelength modes.

The lagging and memory effects that we have described to this point were already present in previous studies of the evolution
of critical flucuations in a spatially uniform cooling plasma~\cite{Berdnikov:1999ph,Mukherjee:2015swa}. 
Our results at the representative value of $r$, $r=1~\fm$, that
we have chosen in plotting Figs.~\ref{fig:phi-fix-r} and ~\ref{fig:phi-fix-r-diff-Gamma}  share some of
the same qualitative features as theirs.
The two important distinctions in the present study are that: (i) 
we are considering a finite, inhomogeneous, fireball undergoing
radial expansion and flow; and (ii) our Hydro+ treatment incorporates the feedback of the critical fluctuations on the hydrodynamic variables. We shall illustrate (i) here, and present results
that bear upon (ii) in Section~\ref{sec:bulk-results}.

%
%
%
 %
%
%

In Fig.~\ref{fig:phi-fix-Q},
we show $\phi(Q)$ at two values of the wave vector $Q$, 
$Q=0.4~\fm^{-1}$ 
and $Q=1.5~\fm^{-1}$, as functions of $r$ at three values of $\tau$.
We show results from Hydro+ calculations done with
two values of $\G_0$, namely $\G_{0}=1~\fm^{-1}$ and $\G_{0}=0.25~\fm^{-1}$.
In all four panels, we compare $\phi(Q)$ with what it
would have been in equilibrium, $\phieq(Q)$.  We
see that the out-of-equilibrium effects 
are much larger at the smaller value of $\G_0$ and 
we see that the magnitude of $\phi(Q)$ as well
as the magnitude of the out-of-equilibrium effects
are much larger at the smaller value of $Q$.
In the top-right panel, where $\Gamma_0$ 
and $Q$ are both larger, the solid curves $\phi(Q)$
are close to the dashed curves $\phieq(Q)$ at all $r$ and
at all three values of $\tau$. We see the largest out-of-equilibrium effects in the bottom-left panel,
where $\Gamma_0$ and $Q$ are both smaller. 

We see two different effects in Fig.~\ref{fig:phi-fix-Q}.
First, we see the dynamics of $\phi(Q)$ lagging behind
those of $\phieq(Q)$, remembering where it used to be.
This is as we have discussed above, but now we
can see the spatial dependence of these phenomena
that are direct consequences of critical slowing down.
The second effect is visible in all but the top-right panel
of Fig.~\ref{fig:phi-fix-Q}, but is most dramatic in the bottom-left panel in which we have chosen a smaller $Q$ to focus on longer wavelength fluctuations and where
we have chosen a smaller $\G_0$ to emphasize out-of-equilibrium effects.
We see in this panel that the peak in $\phi(Q)$ that
is present in the initial conditions that we employ 
in the range of $r$ where the temperature of the
fluid is in the critical regime when we initialize our model
(recall that we assume in our model that $\phi(Q)=\phieq(Q)$
when we initialize the dynamics at $\tau=1~\fm$) moves outwards
at later time.  This shows that the critical fluctuations 
can be carried outwards by advection by the outward 
radial flow of the bulk hydrodynamic fluid.
We can only see this phenomenon, which has
not been reported before, if we pick a small
enough $\G_0$, since otherwise $\phi(Q)$ in this range of $r$ 
relaxes before there is time for it to be carried outwards.
This is a particularly nice illustration of the dynamics incorporated within the equations of Hydro+, and shows that including 
spatial inhomogeneity as well as radial flow is not only
necessary but is also interesting and important. We note, however, that the specific consequences of the advection of
critical fluctuations by the flowing fluid will likely
differ from those we have illustrated in our particular
model study of Hydro+ in action because the specific consequence that we have shown is an artifact of our assumption that
$\phi(Q)=\phieq(Q)$ initially.  
Given the lumpiness of the initial conditions in 
realistic modelling of heavy ion collisions, however,
we expect that in future Hydro+ modelling that is more realistic than ours advection will play some role, although we believe that
the out-of-equilibrium effects originating from $\phi(Q)$ lagging behind $\phieq(Q)$ will be more significant.

We also note that in order to control 
numerical artifacts (fine-scale oscillations in the solid curves) coming from the spatial lattice spacing,
this lattice spacing must be small enough to render the peak in the $\phieq(Q)$ curve smoothly. In our simulations, we have chosen a lattice spacing of $\sim 0.06$~fm, which makes these oscillations almost invisibly small in all the curves we have plotted. Very close inspection shows a 
small trace of these numerical artifacts remaining in the green and blue curves in the bottom-left panel
of Fig.~\ref{fig:phi-fix-Q}, to the left of their peaks.

We close this Section with a speculation about
the value of $\G_0$.  If there really were a critical
point near the $\mu_B=0$ axis, as we have assumed
for this model study, then for such a critical point
it would be reasonable to guess that $\G_0$ is of 
order 1 fm$^{-1}$.
For such 
a critical point, calculations done with $\G_0=0.25~\fm^{-1}$
would not likely be relevant.
However, if there is a critical point in the QCD phase diagram
it is not near $\mu_B=0$. As we discussed in Section~\ref{sec:crit},
in this case the order parameter is a linear combination
of the conserved baryon number density as well as the chiral
condensate, meaning that in this case $\Gamma\propto \G_0 (\xi/\xi_0)^{-3}$
rather than $\Gamma\propto \G_0 (\xi/\xi_0)^{-2}$ as in our
model calculation.
This suggests that if we wish to use Hydro+ calculations done
within our model
to gain qualitative insights into out-of-equilibrium dynamics
near a possible QCD critical point we should use a value
of $\G_0$ that appears unrealistically small in our model.

\subsection{Feedback of critical fluctuations on bulk evolution: qualitative discussion and quantitative results}
\label{sec:bulk-results}

We turn now to what can be seen as the second half
of our study of Hydro+ in action.  In Section~\ref{sec:phi-ev}
we have focused on the out-of-equilibrium dynamics of $\phi(Q)$.
The critical fluctuations described by $\phi(Q)$ are driven
out of equilibrium by the time-dependence of the bulk evolution described by hydrodynamics, with the effects enhanced by
critical slowing down as we have seen.
The second half of the Hydro+ story is that the 
out-of-equilibrium critical fluctuations 
feedback on, and influence, the equation of state and hence
the dynamics of the
ordinary hydrodynamic variables that describe the 
bulk evolution.  We turn now to illustrating these effects.

%
%
%
\begin{figure}
\center
%

\includegraphics[width=0.45\textwidth]{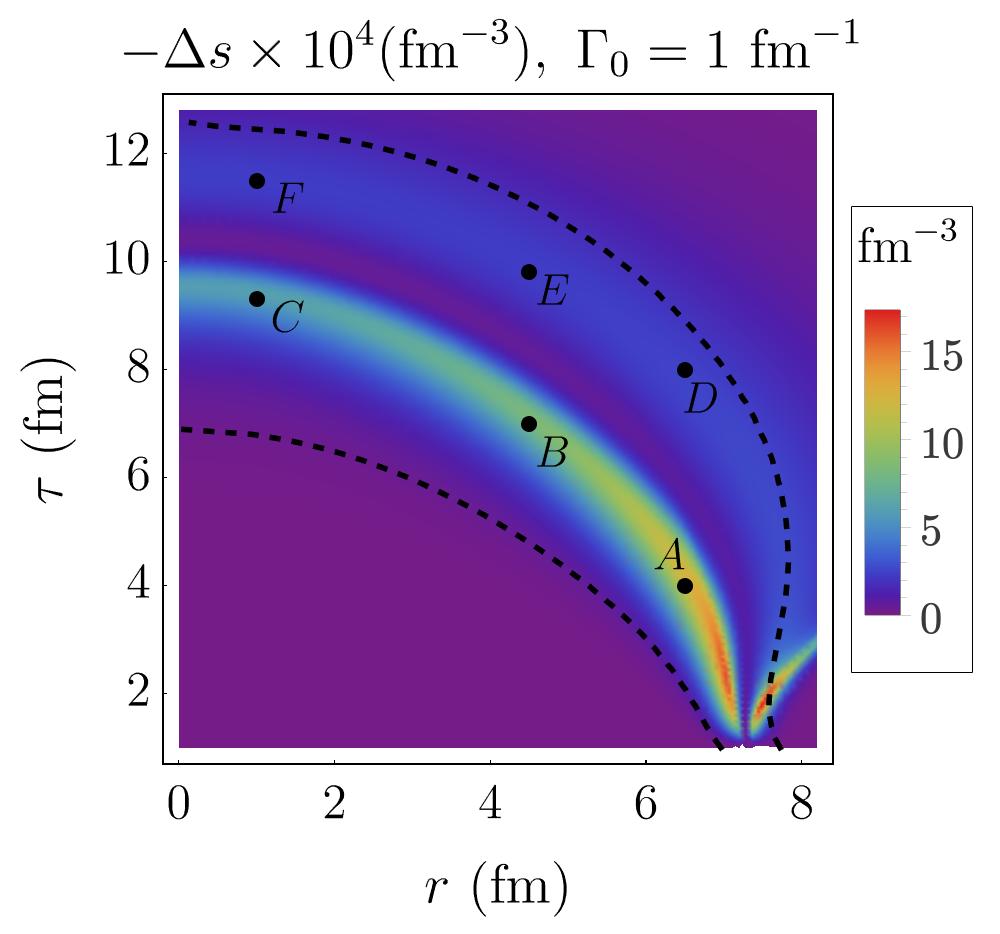}
\includegraphics[width=0.45\textwidth]{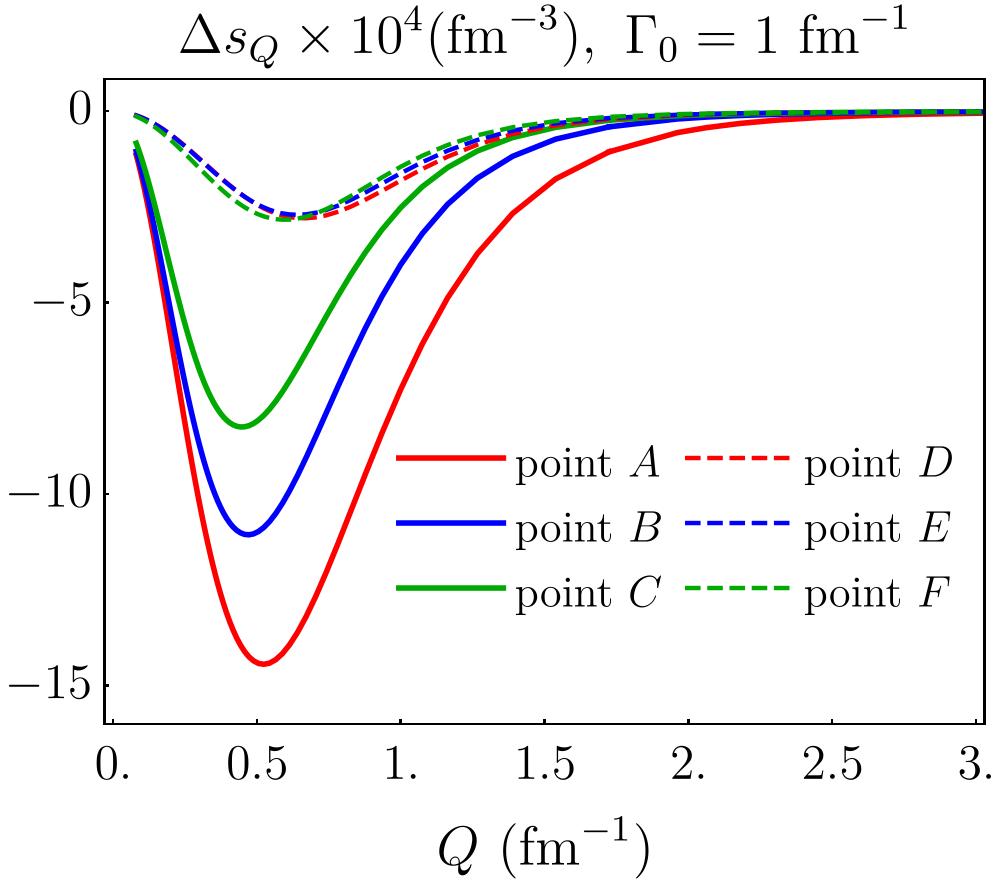}
%
%
\caption{
\label{fig:Deltas}
Left panel: 
$\Delta s$, the difference between the equilibrium entropy 
density $s$ and the Hydro+ entropy density $s_{\plus}$ that includes contributions from the out-of-equilibrium dynamics of $\phi(Q)$. $-\Delta s \equiv s-s_{\plus}$ is plotted vs $(r,\tau)$ 
for $\Gamma_{0}=1~\fm^{-1}$. 
$\Delta s$ is the most direct quantitative measure of 
the influence of out-of-equilibrium critical fluctuations on the equation of state and consequently on the bulk hydrodynamics.
Right panel: 
the integrand in the definition of $\Delta s$, \eqref{Delta-s}, plotted vs.~the wave vector $Q$ at six representative points in the $(r,\tau)$ plane, shown as black dots in the left panel. 
}
\end{figure}
 %
%
%

A key ingredient in the Hydro+ formulation is the 
generalized entropy $s_{\plus}$, or its difference relative to the equilibrium entropy density $\Delta s \equiv s_{\plus}-s$.
$\Delta s$ describes the modification of the entropy density
due to the out-of-equilibrium critical fluctuations described
by $\phi(Q)$. It is 
given explicitly by \eqref{Delta-s}.
As we have discussed in Section~\ref{sec:hydro-plus-review},
from $s_{\plus}$ we can obtain the modified equation of state, namely the modified
pressure $p_{\plus}$ which influences the evolution of radial flow.
In the left panel of Fig.~\ref{fig:Deltas}, 
we plot $-\Delta s$ vs $(r,\tau)$.
To evaluate $\Delta s$ using \eqref{Delta-s}, we have taken our results for $\phi(Q)$ that we obtained by solving the 
Hydro+ equations with $\Gamma_{0}=1~\fm^{-1}$.
We have seen in Fig.~\ref{fig:phi-fix-r} that the evolution of $\phi(Q)$ goes out of equilibrium in two characteristic stages.
First, at earlier times as 
$\phieq(Q)$ rises as the cooling plasma approaches $T_c$ from above
$\phi(Q)$ lags below $\phieq(Q)$.
At later times, as $\phieq(Q)$ drops as the plasma
cools away from $T_c$ toward lower temperature
$\phi(Q)$ lags above $\phieq(Q)$.
We therefore see two bands in the left panel of
Fig.~\ref{fig:Deltas} where $-\Delta s$ is significant
in magnitude.  Between these bands, $T$ passes through $T_c$
and, at a slightly later time, $\phi(Q)$ crosses from below $\phieq(Q)$ 
We see from the figure that $-\Delta s$ 
is larger, meaning that out-of-equilibrium effects are 
larger, in the lower (hotter) of the two bands, where $T$ is approaching $T_c$ from above. This is because here 
the equilibrium $\phieq(Q)$ is shooting upwards
and can get quite far above $\phi(Q)$ whereas in the 
upper (colder) band where $\phi(Q)$ is chasing a falling $\phieq(Q)$ 
there is less separation since $\phieq(Q)$ never drops into
negative territory.
We also see from the figure that $-\Delta s$ is larger at 
larger $r$ in the lower of the two bands. 
This is because the fluid at large $r$ enters the critical regime earlier in time, when the expansion rate is larger. 
In addition, we see the effect of the initial $\phi(Q)$ peak advecting outwards, which creates
the bright band with positive slope in the bottom right corner of Fig.~\ref{fig:Deltas}.
Indeed, the slope of this bright band is very well approximated by the inverse of the local radial fluid velocity, $v_r^{-1}$, of the fluid at its location in $r$ and $\tau$.

$\Delta s$ in \eqref{Delta-s} is given by an integration over all wave vectors $Q$. It  
is instructive to ask which range of $Q$ makes the most important contribution to $\Delta s$. 
To answer this question, 
we pick three representative points in the $(r,\tau)$ plane, 
labelled A, B and C in the left panel of Fig.~\ref{fig:Deltas}
that lie in the lower band where $|\Delta s|$ is significant
and three representative points labelled D, E and F that lie
in the upper band,
and in the right panel of Fig.~\ref{fig:Deltas} we 
plot the integrand in the expression \eqref{Delta-s} for $\Delta s$ 
vs.~$Q$ at all six of these points.
Modes with a large wave vector $Q$ contribute less to the
integrand because these modes remain close to their equilibrium
values.  It is the longer wavelength modes with smaller $Q$ that
are driven farther out-of-equilibrium.
On the other hand, the contribution of modes
with small wave vectors $Q$ to the integrand is
suppressed just by the smaller phase space volume.
So, we see that for each of the three curves
in the right panel of Fig.~\ref{fig:Deltas} 
the integrand in the expression \eqref{Delta-s}
is dominated by wave vectors in a range centered
around $Q = 0.4-0.7~\fm^{-1}$ in our
Hydro+ calculation with $\G_0=1~\fm^{-1}$.  
The modes in this range of wave vectors, which
emerges from the dynamical calculation, make
the most important contribution to the feedback
of the critical fluctuations on the bulk hydrodynamic variables.
This is why
we have chosen $Q=0.4~\fm^{-1}$ as one of the
two wave vectors at which we plotted $\phi(Q)$
in Fig.~\ref{fig:phi-fix-Q}.

We learn an important qualitative lesson from
the right panel of Fig.~\ref{fig:Deltas} that we have taken
advantage of in our calculation, as we describe in Appendix~\ref{sec:coding}.
The dynamical phenomena described by 
Hydro+ come with a natural 
UV cut-off in the integration of $\Delta s$. 
This is the reason why need 
not extend our calculations to modes with
arbitrarily high wave vectors
in practice,
and can focus on the evolution of modes that are
out-of-equilibrium.

%
%
%
\begin{figure}
\center
%
%
%
\includegraphics[width=0.45\textwidth]{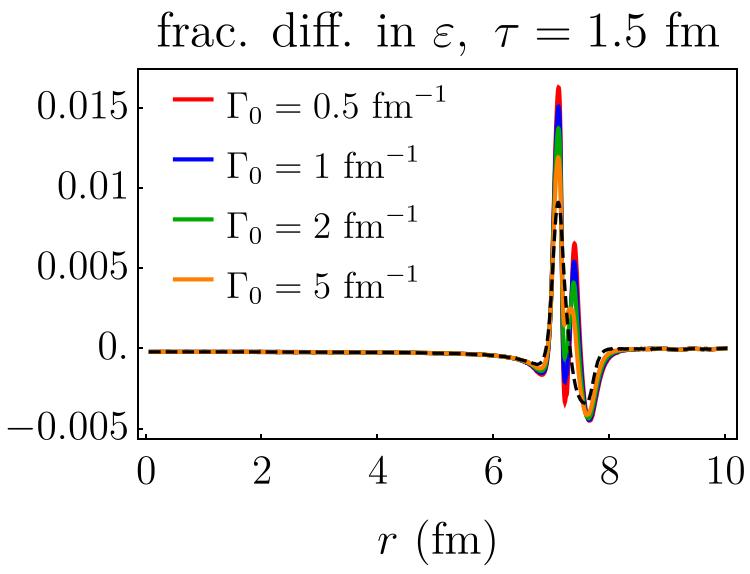}
\includegraphics[width=0.45\textwidth]{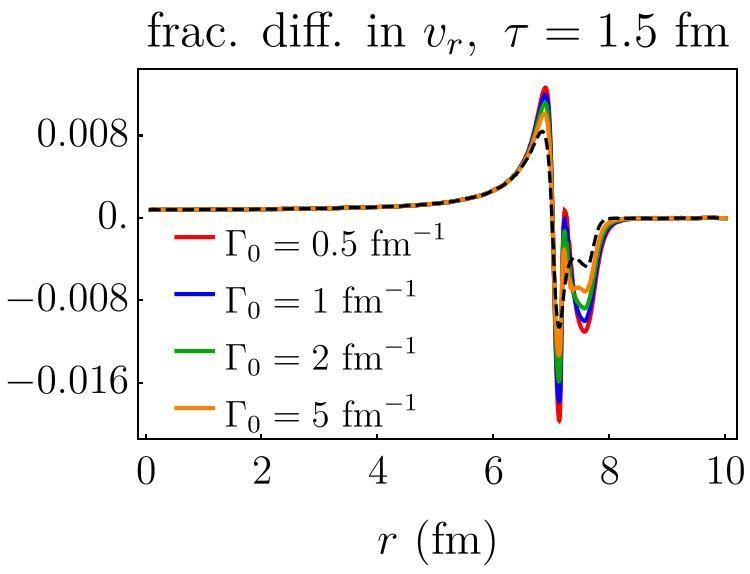}
%
%
\includegraphics[width=0.45\textwidth]{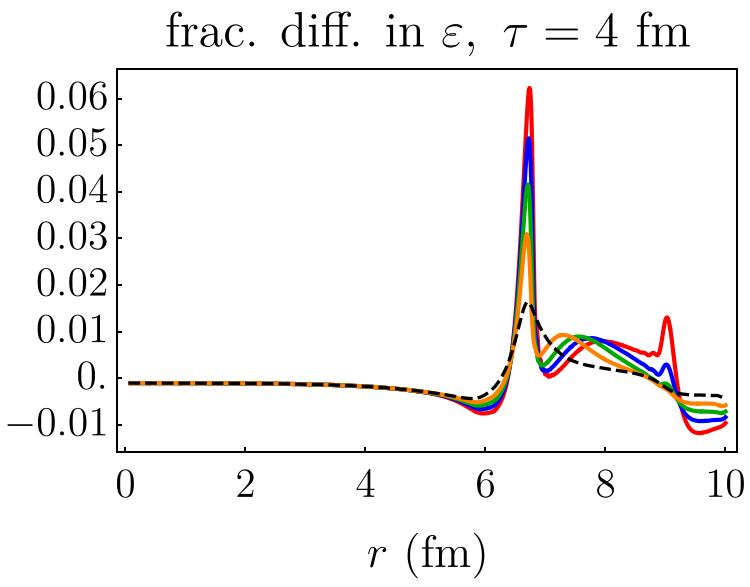}
\includegraphics[width=0.45\textwidth]{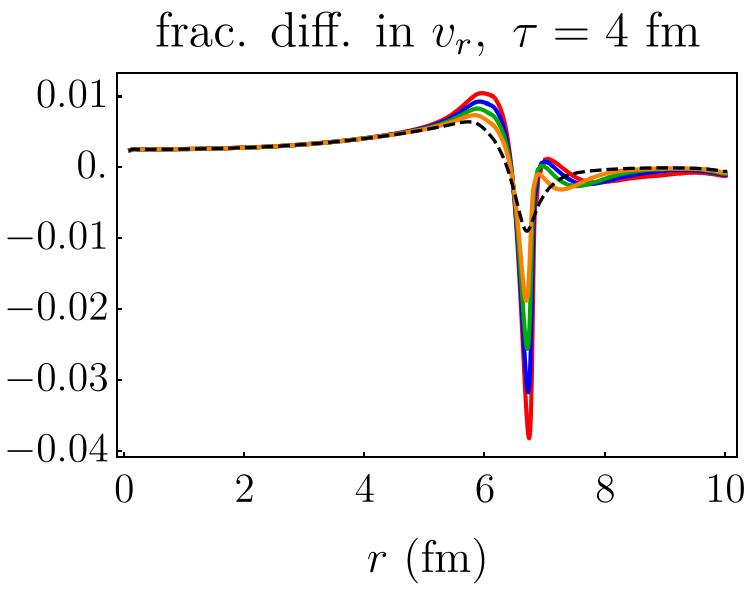}
%
%
\includegraphics[width=0.45\textwidth]{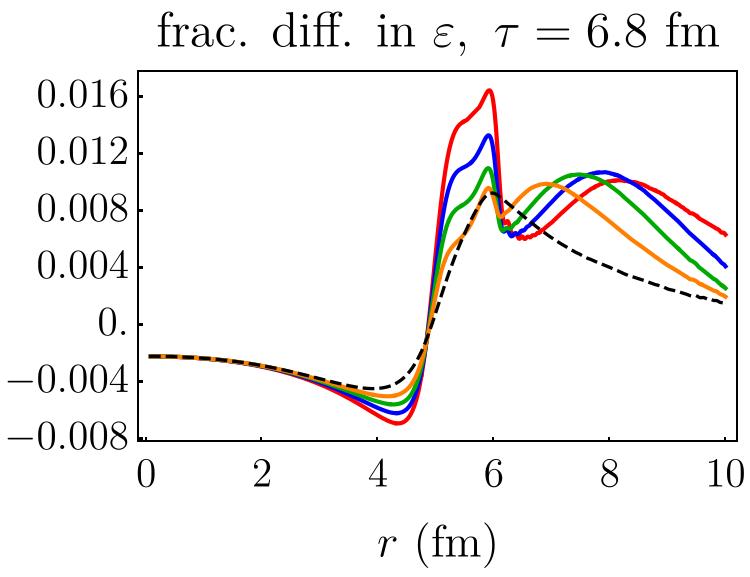}
\includegraphics[width=0.45\textwidth]{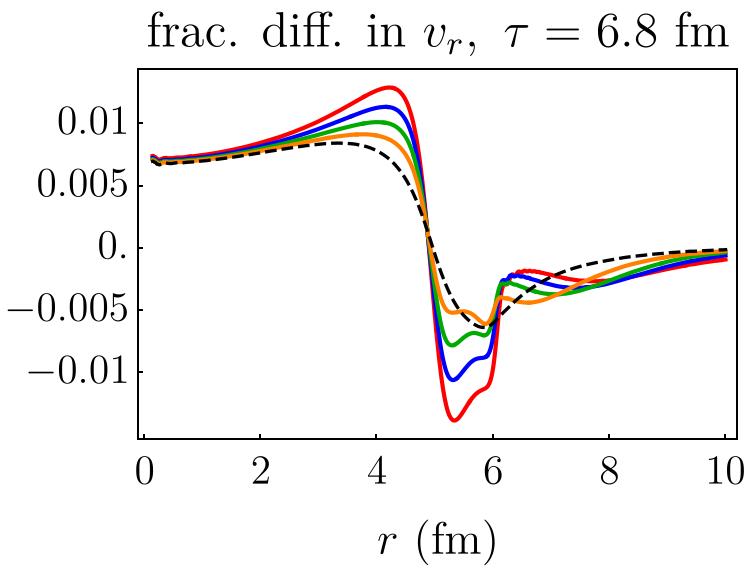}
\caption{
\label{fig:backreaction-v2}
The fractional shift in the 
energy density $\varepsilon$ (left column) and 
radial flow $v_r$ (right column) relative to 
their values in Fig.~\ref{fig:bulk} 
that occurs due to including the effects of: (1) adding the critical point to the equilibrium equation of state (black dashed), and (2) adding, in addition, the back-reaction on the hydrodynamic 
variables coming from the out-of-equilibrium critical fluctuations, as described by Hydro+ (red, blue, green, and orange curves). 
The colored curves are obtained from Hydro+ calculations with three different values of the relaxation rate parameter $\Gamma_0$. 
From top to bottom, the rows correspond to choosing $\tau=1.5$, 4 and 6.8~fm. 
}
\end{figure}
 %
%
%

We are finally ready to look at the
effects on the bulk evolution of the energy density and
radial flow caused by the feedback from the 
out-of-equilibrium dynamics of the critical fluctuations,
as described by Hydro+.  From the $\Delta s$ and $s_{\plus}$ that we have just described we use thermodynamic relations to
obtain the equation of state $p_{\plus}$ whose gradients
drive the radial flow velocity $v_r$, and in Hydro+ as in
ordinary hydrodynamics this radial
flow (together with the boost-invariant longitudinal expansion)
determines how the droplet of plasma cools and how its energy density $\varepsilon$ drops.
In Fig.~\ref{fig:backreaction-v2} we plot results showing how
the dynamics of $\varepsilon$ and $v_r$ in our Hydro+ calculations
differ from those in our starting point, Fig.~\ref{fig:bulk}.
Recall that in Fig.~\ref{fig:bulk}
we solved ordinary hydrodynamic equations, no Hydro+, and
used only the non-critical equation of state from our model.
We shall make comparisons between three cases:
\begin{enumerate}
    \item {\textbf{"B.R":}} solving full Hydro+, i.e. the back-reaction from the slow evolution of $\phi(Q)$ onto
    the evolution of the energy density $\varepsilon$ and radial 
    flow $v_r$  is taken into account.
    \item {\textbf{"no B.R.:"}} solving hydrodynamic equations, no Hydro+, using the critical EoS. 
    \item{\textbf{"no C.P."}} solving hydrodynamic equations using only the non-critical EoS, as in Fig.~\ref{fig:bulk}. 
\end{enumerate}
As we have already noted in Fig.~\ref{fig:EoS}, the difference in the sound velocity between our
critical EoS and our non-critical EoS is 
small, because we have assumed that $\xi/\xi_0$ peaks at 3, rather
than diverging.
Therefore the difference in the bulk evolution for the EoS with and without a critical point is also small. 
For this reason, 
we shall plot the differences:
\bes
\label{Delta-e-vr-def}
\begin{eqnarray}
\Delta \varepsilon_{\noBR}\equiv \varepsilon_{\noBR} - \varepsilon_{\noCP}\, , 
\qquad
\Delta \varepsilon_{\BR}\equiv \varepsilon_{\BR} - \varepsilon_{\noCP}\, , 
\\
\Delta v^{r}_{\noBR}\equiv v^{r}_{\noBR} - v^{r}_{\noCP}\, , 
\qquad
\Delta v^{r}_{\BR}\equiv v^{r}_{\BR} - v^{r}_{\noCP}\, .
\end{eqnarray}
\ees 
$\Delta \varepsilon_{\noBR}$ and $\Delta v^{r}_{\noBR}$ describe the modification of the bulk evolution due to the growth of critical fluctuations in our model as if they were in equilibrium.
The black dashed curves in Fig.~\ref{fig:backreaction-v2} show the fractional differences $\Delta \varepsilon_{\noBR}/\varepsilon_{\noCP}$  and $\Delta v^r_{\noBR}/v^r_{\noCP}$ from \eqref{Delta-e-vr-def}.
Finally, $\Delta \varepsilon_{\BR}$ and $\Delta v^{r}_{\BR}$  
describe the modification of the bulk evolution in our
model due to the growth of critical fluctuations 
as would actually occur, namely out-of-equilibrium, as described
by Hydro+.
The colored curves in Fig.~\ref{fig:backreaction-v2} show the fractional differences $\Delta \varepsilon_{\BR}/\varepsilon_{\noCP}$ and $\Delta v^r_{\BR}/v^r_{\noCP}$ from \eqref{Delta-e-vr-def}. 

Figure \ref{fig:backreaction-v2} showcases the relative contribution of the cases labeled ``B.R.'' and ``no B.R.'' 
relative to the bulk evolution with no critical point. We see 
that for reasonable values of the relaxation rate $\Gamma_0$ in the range  $0.5-2 ~\fm^{-1}$, the effects of 
the back-reaction from the out-of-equilibrium critical fluctuations
are greatest at $\tau\approx4~\fm$ and $r\approx 7~\fm$, where they amount to a $1.5 - 4.5\%$ difference in $\varepsilon$ on top of the $\sim 1.5$\% difference arising solely from including the critical point in the equation of state. 
We have also included results with $\Gamma_0=5~\fm^{-1}$ to
show that at larger values of $\Gamma_0$, where the critical fluctuations track more closely to their equilibrium behavior,
the back-reaction on the hydrodynamic variables coming
from the out-of-equilibrium fluctuations is indeed smaller.

We can understand the qualitative shapes of the 
back-reaction effects depicted by the colored curves
in Fig.~\ref{fig:backreaction-v2} relative to the black dashed ``no B.R.'' curves. 
The largest contribution to the back-reaction 
on the hydrodynamic variables arises
from the modification to the pressure gradient $d(p_{(+)}-p)/dr$, which is given by first and second derivatives of $\Delta s$, see \eqref{Delta-p}. 
Since pressure gradients drive the radial flow,
this translates directly into a modification of $v_r$.
Modifications of $v_r$ yield modifications of the energy
density $\varepsilon$ since $\varepsilon$ is diluted more rapidly
in places where $v_r$ is greater.
Looking at a fixed $\tau$ slice of Fig.~\ref{fig:Deltas}, we see that at some fixed $\tau$ the quantity $-\Delta s$ should have two positive peaks, meaning that its second derivative should change signs four times.  The distortions in the colored ``no B.R.'' energy density curves 
indeed changes sign four times. This is most easily visible in
the middle panels at $\tau=4$~fm where we can see that
(as a function of increasing $r$) the energy density $\varepsilon$
is first below the no-back-reaction black dashed curve, then above, then below, then above and finally 
below again. This is the case also in the upper panels; in the lower panels, the last sign change occurs at larger values of $r$ than we have plotted.
We also note that whenever the radial velocity distortion is negative, the energy distortion is positive, reflecting the fact that when a portion of the fluid expands relatively slower, it cools less, so its energy density decreases less. For example, the largest distortion
in $v_r$ that we see, the downward-going spike at $r$ just below 7 fm
at $\tau=4$ fm, corresponds directly to the largest distortion in $\varepsilon$ which is indeed an upward-going spike.

We close by noting that all the effects of the back-reaction of the out-of-equilibrium
critical fluctuations on the hydrodynamic variables that we
have illustrated in Fig.~\ref{fig:backreaction-v2}
are sufficiently small that in future more realistic modelling of heavy ion collisions it may suffice for phenomenological purposes
to use Hydro+ to describe the growth and out-of-equilibrium dynamics of $\phi(Q)$ while neglecting the back-reaction of
the critical fluctuations on the hydrodynamic variables and
the bulk evolution.

\section{Discussion and Outlook}
\label{sec:conclusion}

We have successfully exercised the newly developed Hydro+ formalism,
testing all of its elements in a concrete model setting where we have
been able to see all aspects of the formalism in action. The model that we have used is simplified in a number of ways, meaning that results from this study
itself cannot be used to compare to experimental measurements in heavy ion collisions, for
example from the Beam Energy Scan program at RHIC.
This means that there are opportunities for future extensions of this work, as we discuss below.  First, though, some remarks about our results are in order, as at a qualitative level they should provide some guidance as to what phenomena to expect in heavy ion collisions that cool near
a critical point in the QCD phase diagram and they should certainly
provide qualitative guidance for future quantitative calculations
of hydrodynamic evolution near a critical point.

Hydro+ allows us to evolve bulk hydrodynamic variables that describe the expansion and cooling of a droplet of QGP together with
critical fluctuations that are necessarily
out-of-equilibrium in a self-consistent fashion, including the
feedback of each upon the other.  In our model we have
seen this interplay manifest explicitly.

As the droplet in our model cools and approaches near
a critical point, we see the magnitude of the critical fluctuations
grow, but lag behind how they would have grown if they were in
equilibrium.  As the critical point is passed and the droplet
cools further, the critical fluctuations decrease in magnitude,
but here again they lag -- staying larger than they would have
if they were in equilibrium.
This lag is the central reason why out-of-equilibrium physics
is important near a critical point, as has long been
realized from more simplified treatments.
What we have been able to do for the first
time is to watch all of this physics
as a function of position in a spatially inhomogeneous model.
We have also seen that the fact that the droplet expands radially
introduces a further phenomenon: once critical fluctuations have
been sourced, they are advected by the flowing fluid, carried outwards to larger radii than where they were born.

A considerable fraction of the technical challenge in implementing Hydro+ comes from calculating the back-reaction that the 
out-of-equilibrium critical fluctuations exert on the 
hydrodynamic variables.  We have completed this computation
explicitly, calculating the back-reaction on the entropy density,
pressure, energy density, speed of sound and radial velocity, and plotting the changes induced in the entropy and energy density as well as in the 
radial velocity.  We find that the back-reaction effects are small, in a few cases larger than 1\% but often smaller than that.  As the model is improved (see below) this should be rechecked but this gives us hope that in future phenenological modelling of heavy
ion collisions at BES energies it may suffice
to use only the ``first half'' of Hydro+, namely the calculation
of the evolution of critical fluctuations in a dynamical hydrodynamic background, without including the ``second half'', namely the
calculation of the back-reaction on the hydrodynamics.
Note also that, as we explained in Section~\ref{sec:EoS},
while the critical fluctuations from a single order parameter
degree of freedom are enhanced near a critical point, the thermodynamics of the bulk comes from a strongly coupled liquid built from 16 bosonic degrees of freedom and 36 fermionic degrees of freedom. Therefore
the influence of critical fluctuations on the entropy density
is small if the fluctuations are in equilibrium, as we saw
in Fig.~\ref{fig:EoS}, and the effects of the back-reaction
from the out-of-equilibrium fluctuations on the entropy density
which are also small can become comparable in magnitude.
The smallness of the effects of the fluctuations on 
the thermodynamic variables indicates that the observables
that will be important in the 
experimental search for a possible critical point
are those sensitive to the direct effects of the fluctuations
themselves, rather than modifications to flow observables.

We close by listing some of the many ways in which our simplified model
can be extended.
\begin{itemize}
    \item 
Developing Hydro+ calculations for hydrodynamic backgrounds
whose expansion is not boost invariant and not azimuthally symmetric.
\item
Developing Hydro+ calculations in which the critical point
is far from $\mu_B=0$, built upon an equilibrium equation
of state like that in Ref.~\cite{Parotto:2018pwx} rather than the simplified
one that we have employed.  Working at nonzero $\mu_B$ introduces the physics of baryon
number diffusion, which will lag due to the enhancement 
in the baryon susceptibility. It may also enhance some
of the effects that we have seen due to the
fact that in this case the order parameter includes
a conserved component the relaxation time will be
somewhat longer than in our calculation.  For this reason,
the back-reaction of the fluctuations on the hydrodynamics
should be analyzed as we have done, but if it is as small
as we have found it may suffice to neglect
the back-reaction in phenomenological modelling, making it possible to rely upon 
hydrodynamic modelling
as is already being 
developed in Refs.~\cite{Denicol:2018wdp,Du:2019obx}, together with a Hydro+ analysis of
the critical fluctuations without back-reaction.
\item
Quantitative calculations of how the Hydro+ fields, including
$\phi(Q)$ which describes the magnitude of critical fluctuations (for example in
in the mass of the proton)
particlize and freeze out are a necessity before phenomenological modelling of experimental observables that
are sensitive to the fluctuations can begin. 
\item
Much work remains to be done in order to model the initial
conditions for Hydro+.  Much work is already underway on
modelling the initial conditions for hydrodynamics~\cite{Shen:2017bsr},
and one can envision future calculations in which standard (non-critical) event-by-event fluctuations in the initial conditions for the energy density and hydrodynamic flow are implemented in an event-by-event Hydro+ calculation.
In addition,
our assumption that $\phi(Q)$ begins in equilibrium must be revisited.
This is perhaps a reasonable assumption deep within the fireball, where
the initial temperatures are hot and initializing $\phi(Q)$ in equilibrium means initializing it at a small value and watching
it grow as the QGP cools, as we have done.  Our strict assumption
of initial equilibrium for $\phi(Q)$ everywhere, 
though, means that we have initialized it
with a large magnitude in an outer shell of radii where the 
temperatures are near the critical point initially. This 
is unrealistic and needs to be handled in a more sophisticated
fashion in future phenomenological modelling.
\end{itemize}
Opportunities abound; we look forward to seeing Hydro+ in action in more and more realistic model settings in anticipation of the day when
predictions from a simulation framework for BES collisions that incorporates a Hydro+ analysis of critical fluctuations can be
compared to experimental data, with the goal of first finding
a critical point in the QCD phase diagram or excluding its
presence in the regime explored in BES energy collisions and second, if one is found, using the comparison between theory and experiment to learn about the out-of-equilibrium dynamics around it.

\begin{acknowledgments}
We are grateful to Marcus Bluhm, Lipei Du, Ulrich Heinz, Iurii Karpenko, Volker Koch, Marlene Nahrgang, Paul Romatschke, Thomas Sch\"afer, Chun Shen and Misha Stephanov for helpful conversations.  
KR gratefully acknowledges the hospitality of the CERN Theory group. GR is supported by a National Science Foundation Graduate Research Fellowship.
This work was supported by the U.S. Department of Energy, Office of Science, Office of Nuclear Physics, within the framework of the Beam Energy Scan Theory (BEST) Topical Collaboration and grant DE-SC0011090. 
\end{acknowledgments}

\begin{appendix}
\section{Some details regarding numerical implementation}
\label{sec:coding}

It is convenient to regroup quantities into dimensionless combinations for numerical implementation. 
We measure wavectors $Q$ in units of $\xi_{0}$
\begin{eqnarray}
\Qscaled\equiv Q\, \xi_{0}\, . 
\end{eqnarray}
According to \eqref{phi-equil},\eqref{Chi-M} and \eqref{OZ}, we have
\begin{eqnarray}
\phieq_{Q} = \frac{c_{M}}{Q^2+\xi^{-2}}\, . 
\end{eqnarray}
Since $\Delta s$ only depends on
\begin{eqnarray}\label{Small-Y-limit}
Y_{\vQ} &\equiv&\frac{\phi_{\vQ}}{\phieq_Q}-1\  
\end{eqnarray}
 which is dimensionless, 
we do our numerical calculations
in terms of rescaled $\phi$ and $\phieq$ such that
\begin{eqnarray}
\label{phieq-scaled}
\phieq_{\Qscaled}= \frac{1}{\Qscaled^2+A}\, , 
\end{eqnarray}
where for later convenience, we have defined a dimensionless quantity
\begin{eqnarray}
\label{A-def}
A&\equiv& \(\frac{\xi}{\xi_{0}}\)^{-2}\, . 
\end{eqnarray}

We next rewrite our expression \eqref{Delta-beta} for $\Delta \b$ and \eqref{Delta-p} for $d\Delta p$ as
\begin{eqnarray}
\Delta\b&=&
\frac{1}{2 w}\,\int_{\vQ}\, \dot{\phieq}_{\vQ}\,Y_{\vQ}\, ,
\end{eqnarray}
and
\begin{eqnarray}
\label{dDelta-p}
d\Delta p
&=&
\(\frac{\b^{2}\, w_{\plus}}{\b_{\plus}\,c_{V}}-c^{2}_{s}\)d\e+
\frac{w_{\plus}}{2\b_{\plus}w}
\int_{\vQ}\, \[Y_{Q}\ddot{\phieq}_{Q}+\(1-Y_{Q}\)\, \(\dot{\phieq}_{Q}\)^2 \]\, \frac{de}{w}
\no \\
&-&\frac{1}{2\b_{\plus}}\int_{\vQ}\,\[ \(\frac{w_{\plus}}{w}\dot{\phieq}_{Q}\)\(1-Y_{Q}\)- Y_{Q}\]\frac{d\phi_{Q}}{\phi_{Q}}\, .
\end{eqnarray}
Likewise, we can write $\Delta p$ in terms of $\varepsilon$ and $Y_{Q}$, obtaining
\begin{eqnarray}
\(\frac{\b_{\plus}}{\b}\)\frac{d\Delta p}{w}
&=&
\frac{1}{2 s}
\int_{\vQ}\, \[ \(\frac{w_{\plus}}{w}\)Y_{Q}\ddot{\phieq}_{Q}+Y_{Q}\, \dot{\phieq}_{Q} \]\, \frac{d\varepsilon}{w}
\no \\
&-&\frac{1}{2s}\int_{\vQ}\,\[ \(\frac{w_{\plus}}{w}\dot{\phieq}_{Q}\)+\frac{Y_{Q}}{\(1-Y_{Q}\)} \]d\,Y_{Q}\, . 
\end{eqnarray}

Continuing, we define $\dot{\phieq}_{Q}, \Ddot{\phieq}_{Q}$ as:
\bes
\begin{eqnarray}
\dot{\phieq}_{Q}
&=& \frac{w}{\phieq_{Q}}\frac{\pd \phieq_{Q}}{\pd \varepsilon}
=- f_{2}\(Q\xi\)\,\(\frac{w}{A}\, \frac{\pd A}{\pd \varepsilon}\)\, , 
\\
\Ddot{\phieq}_{Q}
&=& \frac{w^2}{\phieq_{Q}}\frac{\pd^2 \phieq_{Q}}{\pd \varepsilon^2}
= 
\[
\(f_{2}(Q\xi)\)^2\(\frac{w}{A}\frac{\pd A}{\pd \varepsilon}\)^2
- f_2(Q\xi)\(\frac{w^2}{A}\frac{\pd^2 A}{\pd \varepsilon^2}\)
\]\, .
\end{eqnarray}
\ees
And, since
\bes
\label{dotphieq-numerics}
\begin{eqnarray}
\frac{w}{A}\, \frac{\pd A}{\pd \varepsilon}
&=& c^2_{s}\, \(\frac{T}{A}\frac{\pd A}{\pd T}\)\, ,
\\
\frac{w^2}{A}\frac{\pd^2 A}{\pd \varepsilon^2}
&=&
c^{4}_s
\[-\(\frac{T}{c_V}\frac{\pd c_{V}}{\pd T}\)\(\frac{T}{A}\frac{\pd A}{\pd T}\)
+\(\frac{T^2}{A}\frac{\pd^2 A}{\pd T^2}\)
\]\, ,
\end{eqnarray}
\ees
we have
\bes
\label{dotphieq-expression}
\begin{eqnarray}
\label{dotphi}
\dot{\phieq}_{Q}
&=& 
- f_{2}\(Q\xi\)\,c^2_{s}\, \(\frac{T}{A}\frac{\pd A}{\pd T}\)\, ,
\\
\Ddot{\phieq}_{Q}
&=& 
c^4_{s}\, 
 \Bigg\{
 \[f_{2}(Q\xi)\]^2\(\frac{T}{A}\frac{\pd A}{\pd T}\)^2
 - \nonumber \\ 
 && f_2(Q\xi)\[-\(\frac{T}{c_V}\frac{\pd c_{V}}{\pd T}\)\(\frac{T}{A}\frac{\pd A}{\pd T}\)
 +\(\frac{T^2}{A}\frac{\pd^2 A}{\pd T^2}\) \]
\Bigg\}\, 
\end{eqnarray}
\ees
Here, we will use the expression \eqref{OZ} for the function $f_2$:
\begin{eqnarray}
f_{2}(Q\xi)
= \frac{1}{1+\frac{\Qscaled^2}{A}}\, .
\end{eqnarray}
Therefore $\dot{\phieq}_{Q}$ and $\Ddot{\phieq}_{Q}$ in \eqref{dotphieq-numerics} only depend on $\Qscaled$, 
We finally have:
\bes
\begin{eqnarray}
\label{Delta-beta-numeric}
\(T\Delta\b\)&=&
\frac{1}{2}\,\frac{1}{\(s\xi^3_{0}\)}\,\int\frac{d\Qscaled}{2\pi^2}\, \Qscaled^2\, \dot{\phieq}_{\Qscaled}\,Y_{\Qscaled}\, ,
\\
\label{Delta-p-numeric}
\frac{d\Delta p}{w}
&=&\frac{1}{2}\,\(\frac{w_{\plus}}{w}\, \frac{\b}{\b_{\plus}}\) \frac{1}{\(s\xi^3_{0}\)}
\,\int\frac{d\Qscaled}{2\pi^2}\,\Qscaled^2\,
\[Y_{Q}\Ddot{\phieq}_{\Qscaled}+\(1-Y_{\Qscaled}\)\, \(\dot{\phieq}_{\Qscaled}\)^2 \]\, \frac{de}{w}
\no \\
&-&\frac{1}{2}\, \(\frac{\b}{\b_{\plus}}\)\,\frac{1}{\(s\xi^3_{0}\)} \int\frac{d\Qscaled}{2\pi^2}\,\Qscaled^2\,\[ \(\frac{w_{\plus}}{w}\dot{\phieq}_{\Qscaled}\)\(1-Y_{\Qscaled}\)- Y_{\Qscaled}\]\frac{d\phi_{\Qscaled}}{\phi_{\Qscaled}}\, . 
\end{eqnarray}
\ees
We use \eqref{Delta-beta-numeric} and \eqref{Delta-p-numeric} in our numerical implementation of Hydro+, with the expressions for $\dot{\phieq}_{\Qscaled}$ and $\Ddot{\phieq}_{\Qscaled}$ given by \eqref{dotphieq-numerics}.

In our numerical implementation of Hydro+, in order to evaluate
the integrals over wave vector $Q$ we must discretize $Q$
and we can only keep a finite number of values of $Q$.
Fortunately, as illustrated in the right panel of Fig.~\ref{fig:Deltas} and discussed in the text there, Hydro+ comes with a natural UV
cut-off.  Very high $Q$ modes do not contribute much to the 
Hydro+ integrals that we wish to evaluate because these modes 
stay close to equilibrium.  So, we will perform integrals
over $Q$ by selecting finitely many values $Q_i$, for example
writing \eqref{Delta-s} as 
\begin{equation}
	s_{(+)}\(T\) = \
	s(T) + \frac12 \sum_i dV(Q_i) \
	\left[ \log \left( \frac{\phi_{Q_i}(\tau_n,r)}{\phieq_{Q_i}(T)} \right) \
	- \frac{\phi_i(\tau_n,r)}{\phieq_{Q_i}(T)}  + 1\right]\, ,
\end{equation}
where $dV(Q_i)$ is the volume element for the $i$'th mode, whose
wave vector is $Q_i$. Due to radial
symmetry, we need only specify the magnitude of our wave vectors. Therefore, the volume element satisfies $dV(Q_i) = \frac{Q_i^2}{2 \pi^2} dQ_i$, accounting for the fact that all modes with wave vectors within a shell of thickness $dQ_i$ and radius $Q_i$ contribute to the above integral equally. As we have noted, we choose to discretize unevenly in wave vector. In fact, what we have found convenient is to 
divide the $Q$-integral into three ranges, over each of which
we discretize evenly in inverse-wave-vector, which is to say evenly
in wavelength, but to choose the spacing between the wavelengths of the modes differently in three ranges. 
We do so motivated by the right panel of Fig.~\ref{fig:Deltas},
which tells us that 
we can choose a coarse spacing of $Q_i$'s for both
the shortest and longest wavelength modes, since neither regime
contributes significantly, while choosing a finer spacing of $Q_i$'s 
in the regime of wave vectors whose contribution 
to the integral is most significant.
Specifically, in our calculation we use $120$
modes with $N_{\rm UV}=10$ of them coarsely spaced at large $Q$
and $N_{\rm IR}=10$ of them coarsely spaced at small $Q$ and the 
rest more finely spaced in between, as follows:
\begin{equation}
	Q_i^{-1} = \begin{cases}
      \frac{R_\text{UV}}{N_\text{UV}} i , & \text{if}\ N_\text{UV} \geq i \geq 1 \\
      \frac{R_\text{IR} - R_\text{UV}}{N_\text{int}} (i-N_\text{UV}) + R_\text{UV}, & \text{if}\  N_\text{UV} + N_\text{int} \geq i > N_\text{UV} \\
      \frac{R_\text{max} - R_\text{IR}}{N_\text{IR}} (i-N_\text{int} - N_\text{UV}) + R_\text{IR}, & \text{otherwise}. \\
    \end{cases}
\end{equation}
where we set $R_{\rm UV}=0.5$~fm, and $R_\text{IR} = 6$~fm, meaning that we have 10 UV modes above $Q=2$~fm$^{-1}$ and 10 IR modes below $Q=(1/6)$~fm
$^{-1}$, and 100 modes between $Q=2$~fm$^{-1}$ and $Q=(1/6)$~fm$^{-1}$. Finally, $R_\text{max}$ is the maximum radius of our box, and is 12.5~fm in this paper. (We have doubled the box size to check convergence.) Finally, we define $dQ_i \equiv Q_i - Q_{i-1}$ for $i>1$ and  $dQ_1\equiv dQ_2$.

\section{Feedback from $\phi$ on $\zeta$ and $c^{2}_s$}
\label{sec:qualitative-discussion}

In this Appendix, 
we will provide a qualitative illustration of how effects
originating from deviations between $\phi_{\vQ}$ and its
equilibrium value modify the bulk viscosity and sound velocity,
in so doing extending the previous analysis of Ref.~\cite{Stephanov:2017ghc}.
Our analytical discussion in this Appendix 
is intended only as illustrative; all of these 
out-of-equilibrium effect are taken
into account via the full, numerical, Hydro+ calculation of $\Delta p$.

In order to pursue this illustration analytically as far
as possible, we shall only treat the case where $\phi_{\vQ}$ 
is close to its equilibrium value. That is, 
recalling the definition \eqref{Small-Y-limit},
we shall consider the limit $Y_{\vQ}\ll 1$.
By construction, $Y_{\vQ}$ vanishes when $\phi_{\vQ}$ is in equilibrium. 

We will first assess the contribution
to the bulk viscosity in the limit \eqref{Small-Y-limit}  
by studying how $\phi_{\vQ}$ 
would react in response to expansion of the medium. 
We  recast the equation of motion \eqref{phi-eqn0} for $\phi_{\vQ}$ into an equation of motion for $Y_{\vQ}$
and keep only terms that are linear in $Y_{\vQ}$, obtaining
\begin{eqnarray}
\label{EOM-YQ}
D\, Y_{Q}&=&
- \Gammaeq_{Q}\,Y_{\vQ}+\dot{\phieq}_{Q}\, \frac{D\varepsilon}{w}\, ,
\end{eqnarray}
where we have introduced the abbreviated notation
\label{dotphi-def}
\begin{eqnarray}
\dot{\phieq}_{Q}\equiv w\,\frac{\pd\log\phieq_{Q}}{\pd\varepsilon}\, . 
\end{eqnarray}
In addition, we have replaced $\Gamma(Q)$ with $\Gammaeq_{Q}$ in \eqref{phi-eqn0}, as we did in Section~\ref{sec:Section2}.

Since the equilibration rate $\Gammaeq_{\vQ}$ 
is an increasing function of $Q$,
modes $\phi(Q)$ with a high enough 
momentum $Q$ will always be able to ``catch
up'' with the changing value of the equilibrium $\phieq(Q)$
as the medium expands, meaning that 
their values of $Y_{\vQ}$ can be estimated by finding the
$Y_{\vQ}$ which makes the RHS of \eqref{EOM-YQ} vanish:
\begin{eqnarray}
\label{YQ-approx}
Y_Q &\approx&
\frac{1}{\Gammaeq_{Q}}\, \dot{\phieq}_{Q}\, \frac{D\varepsilon}{w}
\approx
-\frac{\,\dot{\phieq}_{Q}}{\Gammaeq_{Q}}\,\theta\, ,
\end{eqnarray}
where we have used the hydrodynamic equation $D \varepsilon=-\(\varepsilon+p_{\plus}\)\theta+{\cal O}(\pd^2)\approx -w \theta$, 
and where we have replaced  $p_{\plus}$ with $p$, which is adequate for the desired accuracy of the present analysis.
The expression
\eqref{YQ-approx} implies that 
although as $\Gammaeq_{Q}\rightarrow \infty$ (as happens as $Q\rightarrow\infty$) and $Y_Q$ tends to zero as these high
momentum modes attain their equilibrium values, we see that for large but not
infinite $\Gammaeq_Q$ the leading correction 
to $Y_Q$ is proportional to $\theta$. 

To confirm that the out-of-equilibrium contribution \eqref{YQ-approx}
to $Y_{\vQ}$ in turn yields a contribution to $\Delta p$
that multiplies $\theta$ and hence is in fact a contribution
to the bulk viscosity,
we substitute \eqref{YQ-approx} into \eqref{Delta-s}, \eqref{Delta-beta} and \eqref{Delta-p} and obtain:
\begin{eqnarray}
\label{Delta-p-adia}
\Delta p
&\approx&
\[-\frac{T}{2}\, \int_{|\vQ|\geq \QKZ}
\, \frac{1}{\Gammaeq_{Q}}\, \(\dot{\phieq}_Q\)^{2}\]\theta\, ,
\end{eqnarray}
where we have only kept terms up to linear order in $\theta$
and where
we have introduced a lower limit $\QKZ$ in the integral arising in \eqref{Delta-p-adia} since \eqref{YQ-approx} is only valid for high momentum modes which are near equilibrium. 
We note in passing that
the dominant contribution in \eqref{Delta-p-adia} arises from $\Delta\beta$ in the numerator of \eqref{Delta-p}.
We finally substitute \eqref{Delta-p-adia} into the 
constitutive relation \eqref{Tmunu}, and obtain the contribution
to the stress-energy tensor 
driven by the 
expansion of the medium:
\begin{eqnarray}
\label{Tmunu-theta}
\Delta T^{\mu\nu}
=
-\zeta_{\eff}\, \Delta^{\mu\nu}\,\theta \, ,
\end{eqnarray}
with
\begin{eqnarray}
\label{zeta-eff}
\zeta_{\eff}
&\equiv&
\frac{T}{2}\,\int_{|\vQ|\geq \QKZ}
\, \frac{1}{\Gammaeq_{Q}}\, \(\dot{\phieq}_Q\)^{2}
+ \zeta_{\plus}\, ,
\end{eqnarray}
where the first term is the contribution originating from the
out-of-equilibrium dynamics of $\phi(Q)$ that we have estimated
in this Appendix.
It is evident from \eqref{zeta-eff} that this dynamics 
induces an effective bulk viscosity even if $\zeta_{\plus}$ is zero.

It is worth noting that in the long time limit, or in the limit 
in which the expansion is so slow that modes at all $Q$ satisfy \eqref{YQ-approx},
$\QKZ \to 0$ and we then have from \eqref{zeta-eff}:
\begin{equation}
\label{zeta-crit}
\(\zeta_{\rm eff}-\zeta_{\plus}\)
=\frac{1}{2}\, \int_{|\vQ|\geq  0}
\, \frac{1}{\Gammaeq_{Q}}\, \(\dot{\phieq}_Q\)^{2}\, .
\end{equation}
The expression
\eqref{zeta-crit} describes the contribution from $\phi(Q)$ to the bulk viscosity close to equilibrium; 
this expression has been obtained previously by 
diagrammatic calculations based on mode-mode coupling theory~\cite{onuki2002phase} or by solving the linearized Hydro+ equations~\cite{Stephanov:2017ghc}.

We also note that since $\Gammaeq\to 0$ near the critical point, 
which is the phenomenon of critical slowing down, 
the expression \eqref{zeta-crit} is only valid
near a critical point for {\it very} slow expansion.
That said, we observe that \eqref{zeta-crit} indicates
that when $\Gammaeq \to 0$ the contribution
$\(\zeta_{\rm eff}-\zeta_{\plus}\)$ becomes singular.
It is this observation that motivates us to simplify
our model calculation by choosing $\zeta_{\plus}=0$, 
meaning that the only contributions to the bulk viscosity
are those that come from the out-of-equilibrium dynamics
of $\phi(Q)$ through its contribution to $\Delta p$.
In any realistic context,
$\QKZ\neq 0$ because of critical slowing down.
Consequently, the ratio
\begin{eqnarray}
\frac{\frac{1}{2}\,\int_{|\vQ|\geq \QKZ}
\, \frac{1}{\Gammaeq_{Q}}\, \(\dot{\phieq}_Q\)^{2}
}{\frac{1}{2}\,\int_{|\vQ|\geq 0}
\, \frac{1}{\Gammaeq_{Q}}\, \(\dot{\phieq}_Q\)^{2}
}
\end{eqnarray}
can be of the order unity. 

We turn now to looking at how the out-of-equilibrium 
dynamics of the
low momentum modes of $\phi(Q)$ contribute
to $\Delta p$, and through $\Delta p$ serve to modify
the sound velocity.  In the remainder of this Appendix, we
provide an illustration of this effect.

Let us express $d p_{\plus}$ in terms of $d\varepsilon$ and $d\phi_{\vQ}$ using \eqref{dDelta-p} and $dp=c^{2}_{s}d\varepsilon$, 
and take the limit $Y_{\vQ}\ll 1$:
\begin{eqnarray}
\label{Delta-p-de}
d p_{\plus}
&\approx&
\[c^{2}_{s}+\frac{1}{2 s}
\int_{\vQ}\,  \(\dot{\phieq}_{Q}\)^2 \]d\varepsilon\,
-\frac{1}{2\b}\int_{\vQ}\,\dot{\phieq}_{Q}\,\frac{d\phi_{Q}}{\phi_{Q}}\, . 
\end{eqnarray}
To derive \eqref{Delta-p-de}, 
we also made the replacement $w_{\plus}\to w$ and $\b_{\plus}\to \b$ in \eqref{dDelta-p}. 
We see from the expression
\eqref{Delta-p-de} 
that through its contribution to $\Delta p$ 
the out-of-equilibrium dynamics of $\phi(Q)$
contributes to the square of the sound velocity, $\partial p_{\plus}/\partial\epsilon$, which we shall denote by $c_{s,{\rm eff}}^2$
to distinguish it from the equilibrium $c_s^2$. From \eqref{Delta-p-de}, we have
\begin{eqnarray}
\label{Delta-cs2-approx}
\Delta c^2_{s,\eff}&\equiv&c^2_{s,\eff}-c^2_{s}
\approx
\frac{1}{2 s}\,\int_{|\vQ|\leq \QKZ}\, \(\dot{\phieq}_{Q}\)^2
\no \\
&=&
\frac{c^{4}_{s}}{2 s}\,
\int_{|\vQ|\leq \QKZ}\, \[f_{2}(Q\xi)\]^{2}\,
\(\frac{\xi}{\xi_{0}}\)^{4}\(T\frac{\pd}{\pd T} \( \frac{\xi}{\xi_{0}}\)^{-2}\)^{2}
\no \\
&\approx& \frac{1}{2s}\, \(\frac{\xi}{\xi_{0}}\)^{4}\(\frac{T_{c}}{\Delta T}\)^{2}\,\int_{|\vQ|\leq \QKZ}\, \[f_{2}\(Q\xi\)\]^{2}\, .
\end{eqnarray}
We have introduced an upper bound in the integration, $\QKZ$, 
in order to focus on far-from-equilibrium modes, 
and used \eqref{dotphi} to obtain the second line of \eqref{Delta-cs2-approx}. Note that $f_2$ is the universal
scaling function introduced in \eqref{phi-equil}.
In the third line of \eqref{Delta-cs2-approx},
we used the approximation 
\begin{eqnarray}
\label{dlogxi-dlogT}
\(T\frac{\pd}{\pd T} \( \frac{\xi}{\xi_{0}} \)^{-2} \)^{2} 
\sim \(\frac{T_{c}}{\Delta T}\)^{2}\, ,
\end{eqnarray}
to simplify the expression.
Note that 
\eqref{dlogxi-dlogT} is consistent with our parameterization \eqref{xi-para} 
of $\xi$, which satisfies $\(\xi/\xi_{0}\)^{-2}\sim |T-T_{c}|$ around $T_{c}$. 
Note also that \eqref{Delta-cs2-approx} is positive definite, meaning that the out-of-equilibrium $c_{s,{\rm eff}}^2$ is larger than $c_S^2$,
which is to say the out-of-equilibrium pressure $p_{\plus}$ is stiffer than the equilibrium $p$.

In order to get a qualitative sense of the importance
of the out-of-equilibrium correction to the equation of state,
we close this Appendix by comparing $\Delta c^2_{s,\eff}$ in \eqref{Delta-cs2-approx} to the difference between 
the equilibrium $c^{2}_{s}$ for our non-critical equation
of state and our critical equation of state:
 \begin{eqnarray}
 \label{Delta-cs2-Def}
 \Delta c^{2}_{s}&\equiv&
 \frac{s}{c^{\noCP}_{V}} -\frac{s}{c^{\noCP}_{V}+c^{\crit}_{V}}
\approx
\(\frac{s}{c^{\noCP}_{V}}\)^{2}\, \frac{c^{\crit}_{V}}{s}\, 
\approx
c^{4}_{s} \, \frac{c^{\crit}_{V}}{s}\, ,
 \end{eqnarray}
where we have used the fact $C^{\noCP}_{V}\gg C^{\crit}_{V}$, see Sec.~\ref{sec:EoS}. 
Since $c^{2}_{s}$ is one of the important parameters that controls the hydrodynamical evolution, 
this comparison provides qualitative 
guidance as to whether $\Delta \varepsilon_{\BR}$ ($\Delta v^{r}_{\BR}$) is comparable with $\Delta \varepsilon_{\noBR}$ ($\Delta v^{r}_{\noBR}$) (cf.~\eqref{Delta-e-vr-def}).  
In particular, if $\Delta c^{2}_{s}$ is of the same order as $\Delta c^{2}_{s,\eff}$, 
we also expect $\Delta \varepsilon_{\BR}$ ($\Delta v^{r}_{\BR}$) and $\Delta \varepsilon_{\noBR}$ ($\Delta v^{r}_{\noBR}$) to be similar in magnitude.

The explicit ansatz for $C^{\crit}_{V}$ that we employ in our
model is given by~\eqref{CV-ansatz}; we reproduce it here for convenience
\footnote{In \eqref{CV-ansatz}, $T_{c}/\Delta T$ has been replaced by $5$, the value we have used in our model calculation.}: 
\begin{equation}
\label{CV-ansatz0}
c_V^{\rm crit}(T) = 
\(\frac{T_{c}}{\Delta T}\)^{2}\, C_{M,{\rm Is}}
\end{equation}
where (c.f.~Ref.~\cite{kardar2007statistical})
\begin{eqnarray}
\label{CM-Is0}
C_{M,{\rm Is}}
&=&\frac{1}{2}\, \int_{Q}\, f^{2}_{2}\(Q\xi\)\, \(\frac{\xi}{\xi_{0}}\)^{4}
=\frac{1}{16\pi}\, \xi^{-3}_{0}\, \(\frac{\xi}{\xi_{0}}\)\, , 
\end{eqnarray}
where here $\xi$ and $\xi_0$ are Ising model quantities.
We have used
\begin{eqnarray}
\int^{\infty}_{0} dx\, x^{2} \(\frac{1}{1+x^{2}}\)^{2}=\frac{\pi}{4}\, ,
\end{eqnarray}
to evaluate the integration over $Q$.
Substituting \eqref{CM-Is0} and \eqref{CV-ansatz0} into \eqref{Delta-cs2-Def}, 
we have
\begin{eqnarray}
\label{Delta-cs2-int}
 \Delta c^{2}_{s}
 =\frac{1}{2s}\,\(\frac{\xi}{\xi_{0}}\)^{4}\,  \(\frac{T_{c}}{\Delta T}\)^{2}\,\int_{|\vQ|\leq\infty}\, f^{2}_{2}\(Q\xi\)\, . 
\end{eqnarray}
Taking the ratio between $\Delta c^{2}_{s}$ in \eqref{Delta-cs2-int} with $\Delta c^2_{s,\eff}$ in \eqref{Delta-cs2-approx}, 
we finally have:
\begin{eqnarray}
\label{cs-ratio}
\frac{\Delta c^{2}_{s,\eff}}{\Delta c^{2}_{s}}
&\approx&
\frac{\int_{|\vQ|\leq \QKZ}\, \[f_{2}\(Q\xi\)\]^{2}}{\int_{|\vQ|\leq \infty}\, \[f_{2}\(Q\xi\)\]^{2}}\, . 
\end{eqnarray}
We note the integration over $Q$ in \eqref{Delta-cs2-int} will be saturated when $Q\geq \xi^{-1}$. 
Therefore the ratio in \eqref{cs-ratio} will be of order unity if $Q^{*}\sim \xi^{-1}$, a result that seems quite natural indeed.

\end{appendix}

\clearpage

\bibliographystyle{JHEP}
\normalem
\bibliography{Hydro_Plus_in_Action}

\end{document}